\documentclass[a4paper,12pt,english]{article}

\usepackage{amsfonts,bm,amssymb,euscript,array,babel,cite}
\usepackage[all,color]{xy}
\usepackage{relsize}
\usepackage{tikz}
\usepackage{url}
\usepackage{amsmath,amsthm}
\usepackage[]{graphicx}
\usepackage[T1]{fontenc}
\usepackage{framed}

\usepackage{empheq}

\makeatletter

\newcommand{\dd}{\mathrm{d}}
\newcommand{\dD}{\mathrm{D}}

\newcommand{\w}{\wedge}
\newcommand{\bbm}{\left(\begin{matrix}}
\newcommand{\ebm}{\end{matrix}\right)}
\newcommand{\beq}{\begin{eqnarray}}
\newcommand{\eeq}{\end{eqnarray}}

\makeatother

\newcommand*\circled[1]{\tikz[baseline=(char.base)]{\node[shape=circle,draw,minimum width=1cm,inner sep=2pt] (char) {$#1$};}}

\newtheorem{prop}{Proposition}[section]
\newtheorem{theorem}[prop]{Theorem}
\newtheorem{lemma}[prop]{Lemma}
\newtheorem{defn}[prop]{Definition}

\newcommand{\sfrac}[2]{{\textstyle\frac{#1}{#2}}}

\newcommand{\be}{\begin{equation}}
\newcommand{\ee}{\end{equation}}

\newcommand{\beqa}{\begin{eqnarray}}
\newcommand{\eeqa}{\end{eqnarray}} 
\def\nn{\nonumber} \def \bea{\begin{eqnarray}} \def\eea{\end{eqnarray}}

\newcommand{\barr}{\begin{array}}
\newcommand{\earr}{\end{array}}
\numberwithin{equation}{section}
\newcommand{\mf}{\mathfrak}

\def\a{\alpha}  
 \def\g{\gamma} 
 \def\d{\delta} 

\def\l{\lambda}   
 \def\o{\omega}   
 \def\S{\Sigma}

\def\cA{{\cal A}}   
  \def\cF{{\cal F}} 
\def\cG{{\cal G}}   
  \def\cL{{\cal L}}

  \def\cU{{\cal U}}
  \def\cX{{\cal X}}

\def\R{{\mathbb R}} \def\C{{\mathbb C}}  

\def\Z{{\mathbb Z}} \def\one{\mbox{1 \kern-.59em {\rm l}}}

\def\bit{\begin{itemize}} \def\eit{\end{itemize}}

\def\({\left(} \def\){\right)}

 \allowdisplaybreaks[3]

\begin{document}

\makeatother


\parindent=0cm

\renewcommand{\title}[1]{\vspace{10mm}\noindent{\Large{\bf

#1}}\vspace{8mm}} \newcommand{\authors}[1]{\noindent{\large

#1}\vspace{5mm}} \newcommand{\address}[1]{{\itshape #1\vspace{2mm}}}


\begin{titlepage}

\begin{flushright}
\today 
\end{flushright}

\begin{center}

\vskip 3mm

\title{{\Large Strings in Singular Space-Times and\\ Their Universal Gauge Theory}}


 \authors{
 Athanasios {Chatzistavrakidis}$~^{\sharp,}$$\footnote{Emails: a.chatzistavrakidis$*$rug.nl, andreas.deser$*$itp.uni-hannover.de, larisa$*$irb.hr, strobl$*$math.univ-lyon1.fr.}$, 
Andreas Deser$~^{\star}$\\
 Larisa Jonke$~^{\circ}$
 and Thomas Strobl$~^{\dagger}$
 }

\vskip 1mm

\address{
$^{\sharp}$Van Swinderen Institute for Particle Physics and 
Gravity, University of Groningen, \\ 
Nijenborgh 4, 9747 AG Groningen, The Netherlands

 $^{\star}$
Institut f\"ur Theoretische Physik,  
Leibniz Universit\"at Hannover,\\  Appelstra{\ss}e 2, 30167 Hannover, Germany 

$^{\circ}$ 
Division of Theoretical Physics, 
 Rudjer Bo$\check s$kovi\'c Institute, \\
 Bijeni$\check c$ka 54, 10000  Zagreb, Croatia

$^{\dagger}$
Institut Camille Jordan, Universit\'e Claude Bernard Lyon 1, \\
43 boulevard du 11 novembre 1918, 69622 Villeurbanne cedex, France

}

\smallskip

\end{center}

\vskip 1mm

 \begin{center}
\textbf{Abstract}
\vskip 3mm
\begin{minipage}{14cm}%
We study the propagation of bosonic strings in singular target space-times. For describing this, 
we assume this target space to be the quotient of 
a smooth manifold $M$ by a singular foliation ${\cal F}$ on it. Using the technical tool of a gauge theory, we propose a smooth functional for this scenario, such  that the propagation is assured to lie in the singular target on-shell, i.e.~only after taking into account the gauge invariant content of the theory. One of the main new aspects of our approach is that we do not limit ${\cal F}$ to be 
generated by a group action. 
We will show that, whenever it exists, the above gauging is effectuated by a single geometrical and universal gauge theory, whose target space is the generalized tangent bundle $TM\oplus T^*M$.
\end{minipage}



 \end{center}

\end{titlepage}

\tableofcontents

\section{Introduction}
In its perturbative description, String Theory is described by a two-dimensional sigma model. The target space is the classical geometrical background, formed as an average over the quantum string, satisfying Einstein equations in the leading order so as to guarantee conformal invariance of the worldsheet theory, and serving as the starting point of the perturbative expansion. 

Already classical Einstein gravity is doomed to produce singularities by its own time evolution, all the more we expect them to arise in the quantum world. One of the problems for a classical description of the string in a singular space-time, at least if the singularity prohibits a differential structure, is that the conventional description in terms of a sigma model breaks down. 

\begin{figure}[ht]\label{foliation}
\centering
\includegraphics[width=1\textwidth]{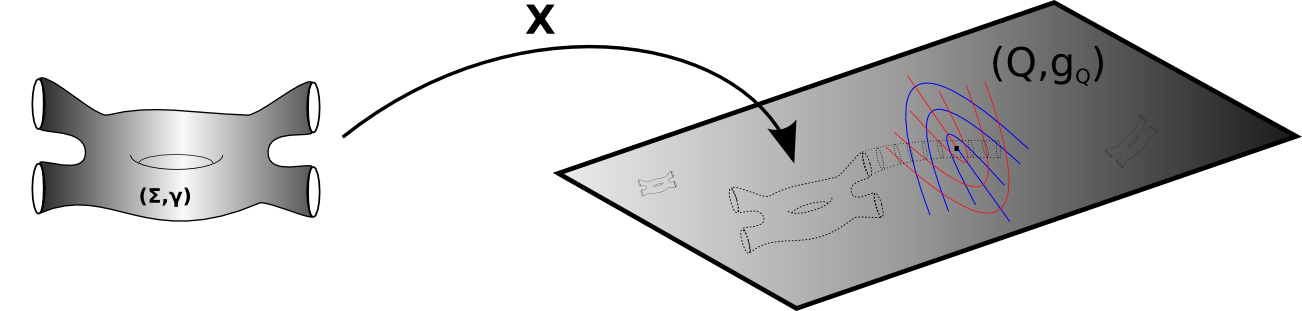}
\caption{String propagation in a quotient $Q=M/{\cal F}$ with ${\cal F}$ being a singular foliation.}
\end{figure}

A large class of singular manifolds $Q$ can be understood as arising as the quotient of a smooth manifold $M$ with respect to a singular foliation ${\cal F}$ on it. In this paper, we will investigate a string propagation in a target $Q$ that arises in this way (see Figure \ref{foliation}). 
The mathematical tool that we will use for its description is the one of gauge theories, albeit in a much more general framework than the standard one using group actions. 

In the subsequent section, we provide the precise setting concerning the singular foliations as well as the type of gauge theories that we will consider. In particular, we will require that the singular foliation is generated by the involutive image of a vector bundle map $\rho \colon E \to TM$. The gauge fields $A$ that we will introduce then are 1-forms on the world-sheet $\Sigma$ with values in the vector bundle $E$. The standard picture of Lie algebra valued 1-forms arises from this for the special case that $E=M\times \mathfrak{g}$, where $\rho$ then corresponds to the action of the Lie algebra $\mathfrak{g}$ on $M$. 

One important objective of Sec.~\ref{Orientation} is to clarify the precise meaning of the gauging of a sigma model. In particular, it should be guaranteed at least locally on $\Sigma$ that the gauge equivalence classes of maps into $M$, where the gauge equivalence precisely corresponds to the foliation ${\cal F}$, corresponds to a free movement of the string on the quotient $Q = M/{\cal F}$. This can actually be tested only for the case that $Q$ is smooth, so that we can compare to a sigma model description with target $Q$. 

Even in the context of conventional gauging of Lie group actions, this is astonishingly not always the case, at least if the background carries a non-trivial $H$-flux, described by a Wess-Zumino term in the sigma model picture. {\em If} the gauged sigma model passes the test of a good quotient description, we call the gauging strict. The G/K WZW model in its conventional description turns out to be a non-strict gauging, for example.

In section \ref{Sec2}, we determine the most general conditions for a metric $g$ and 2-form field $B$ on $M$, such that the gauging can be effectuated by means of minimal coupling, where ordinary derivatives $\dd X$ are replaced by covariant ones $\dD X := \dd X - \rho(A)$. The gauging is always strict in this case. These results are summarized in the propositions \ref{Prop1} and \ref{Prop2} as well as in Theorem \ref{Thmred}.

Section \ref{Sec3} contains the main results of the paper. The gauged sigma models are extended to the case containing an $H$-flux that is not necessarily exact ($H=\dd B$). In the gauging we keep minimal coupling in the $g$-part of the gauged sigma model description, making a general ansatz in the metric independent sector. This gives new conditions on $g$ and $H$ for a gauging to exist, while still far more general than the conventional ones, even in the case that the foliation is generated by a Lie algebra action. 

Most notably, there is a universal form of the gauged theory, depending only on the original geometrical data before the gauging and having as target the generalized tangent bundle $TM \oplus T^*M$. If the gauging exists, i.e.~if the fairly general conditions on $g$ and $H$ in relation to the foliation ${\cal F}$ can be met, this gauged action $S_{gauged}$ can be obtained as a pull-back of a universal action $S_{univ}$. The situation is summarized in the following commutative diagram 
\begin{equation}\label{diagrammain}
  \xymatrix{ & \:\, \circled{\cU} \ar[d]^{{S_{univ}}}
  \\
    \circled{\cA} \ar@{-->}[ur]^{\mathlarger{\exists ! \:\,\widehat\sigma}}   \ar[r]^{{\:\,\, {S_{{gauged}}}}} & \R} 
\end{equation}
where ${\cal A}$ denotes the set of string maps $X \colon \Sigma \to M$ together with the gauge field $A \in \Omega^1(\Sigma, X^*E)$; together this can be also expressed as vector bundle morphisms $a \colon T\Sigma\to E$. Likewise, ${\cal U}$ denotes the set of vector bundle morphisms $u \colon T\Sigma \to TM \oplus T^*M$. The map $\hat{\sigma}$ between these two sets is induced by a bundle map $\sigma  \colon E \to TM \oplus T^*M$ extending the above-mentioned map $\rho$ and having an involutive and isotropic image, when $TM \oplus T^*M$ is viewed upon as an $H$-twisted standard Courant algebroid. For a fixed choice of $S_{gauged}$, the map $\sigma$ and thus also $\hat{\sigma}$, is unique, moreover. The main results of this paper are summarized in the Theorems \ref{Thmmain1}
and \ref{Thmmain2}. 

Section \ref{Examples} illustrates, in its first part, the somewhat abstract results of Sec.~\ref{Sec3} by means of some simple examples.
The main result of this section is a case study of
the WZW model for $G=SU(2)$: we prove explicitly that strict gauging of the adjoint orbits is obstructed in this case. However, we also show that the unintended freezing of the movement of the string on the quotient (which is an interval in this case), can be avoided except for a region of arbitrarily small volume by a different choice of the action. We call this an almost strict gauging then.

Appendix \ref{AppendixA} contains the conditions on 
$g$ and $H$ for the case that one makes an even more general ansatz for the gauged action, relaxing minimal coupling also in the metrical sector.

\section{First Orientation and the Setting}
\label{Orientation}

The basic idea of strings propagating in an $n\geq 4$-dimensional spacetime $M$ is realized, in a first step and on the bosonic level, by a functional $S$ on maps $X$ from a two-dimensional worldsheet manifold $\Sigma$ into the possibly curved target space $M$, cf.~Figure 2.
\begin{figure}[ht]
\centering
\includegraphics[width=1\textwidth]{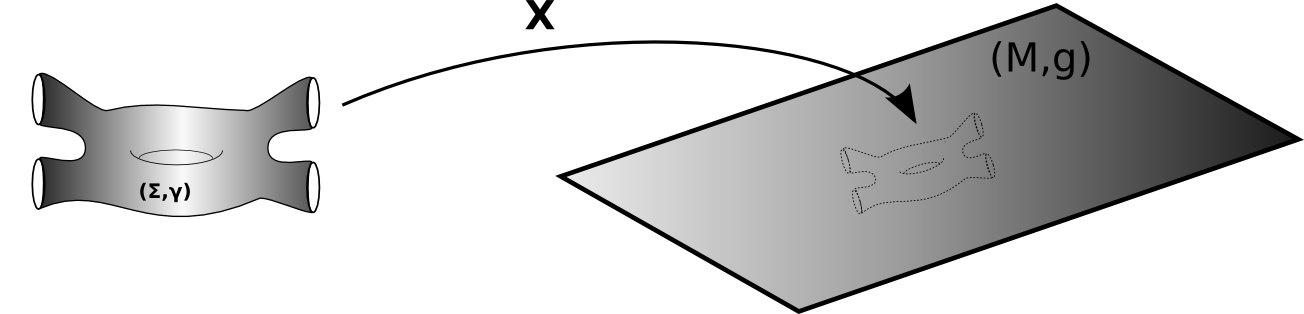}
\caption{The application $X$ maps the world-sheet $\Sigma$ of the string, homeomorphic to a punctured Riemann surface, into the space-time $M$. For a local embedding, the conformal class of $\gamma$ is induced by $g$ on-shell.}
\end{figure}
  In its simplest form this functional is \cite{Polyakov}:
\begin{equation} \label{SPoly}
S[X,\gamma] = \int_\Sigma \sfrac{1}{2} g_{ij}(X) \partial_\mu X^i \partial_\nu X^j \gamma^{\mu \nu} \sqrt{\gamma} \, \dd^2 \sigma  \, .
\end{equation}
Here $X \colon \Sigma \to M$, $\gamma$ is a metric on $\Sigma$ used as an independent variable, and $g$ is the fixed metric (or pseudo-metric) on $M$.\footnote{We use the following notation: if $(x^i)_{i=1}^n$ are local coordinates in $M$, then 
$X^i = X^* x^i$. And $g_{ij}(X) \equiv X^* g_{ij}$ are the pulled-back components
of $g=g_{ij} \dd x^i \otimes \dd x^j\equiv \frac{1}{2} g_{ij} \dd x^i \vee \dd x^j$. We use a similar notation in the 
sequel without extra mention; in addition, we will sometimes simply write $g_{ij}$ for $g_{ij}(X)\equiv X^* g_{ij}$ if
the pullback is understood from the context.}  In a first step  $\gamma$ is often kept fixed and the two-dimensional scalar fields $(X^i)_{i=1}^n$ are quantized according to the standard rules corresponding to the above functional for fixed background $\gamma$; thereafter, the additional equations $T^{\mu \nu} \equiv \frac{\delta S}{\delta \gamma_{\mu \nu}} = 0$ are implemented in one way or another as additional constraints. Full consistency on the quantum level then in fact requires the additional inclusion of fermionic fields and the restriction to a particular dimension $n$ of the target. Also the string is supposed to effectively generate its own target geometry, so in particular also the metric $g$, and non-perturbative corrections are taken into account eventually in form of D-branes (sectors of the theory described perturbatively by particular boundary conditions on $X$ with respect to appropriate submanifolds $D \subset M$ in \eqref{SPoly}). In the spectrum for the string in flat space, one finds beside the ``graviton'' corresponding to the metric $g$ also the degrees of freedom of an antisymmetric 2-tensor on $M$, which then in turn preferably should be used already by a corresponding addition to the original functional \eqref{SPoly}, turning it into 
\begin{equation} \label{SgB}
S_{bos} = \int_\Sigma \sfrac{1}{2}g_{ij} \, \dd X^i \wedge \ast \dd X^j + \int_\Sigma X^* B
\end{equation} 
for a 2-form $B$ on $M$. We used a more condensed notation in \eqref{SgB}; in particular, the metric $\gamma$ is hidden in the Hodge duality operation $\ast$ now.\footnote{We dropped eventual dilaton contributions to the action, since they correspond to one-loop terms; here we focus on leading 
order contributions.
}

We assume all this to be known certainly, cf.~e.g.~the textbooks \cite{Green,Polchinski}, and only focus on the bosonic and classical content of the theory with a fixed worldsheet metric $\gamma$ henceforth. Our goal is to replace the target manifold $M$ by some possibly singular space. One way of achieving this is, e.g., by considering the string to move in the target space \emph{factored} by some group acting on it. Let for example $G=\Z_2$, $M=\R^n$, and the action of the non-trivial element $g \in G$ on $M$ be given by $g 
\cdot x^i = - x^i$ for all $i=1, \ldots ,n$. Here $x^i$ are coordinates on $M$ and the fields $X^i$
on the worldsheet $\Sigma$ result from them by pullback, $X^i =X^*(x^i)$. If we consider only fields $X^i$ such that they are invariant in one way or another with respect to this group action, then we may say that the string is effectively propagating in the cone $\R^n/\Z_2$.
 One may consider doing this for any discrete $G$ acting on $M$ like this \cite{orb1,orb2}. If then the action is not properly discontinuous, the quotient space $M/G$ may have singularities of various kinds. 

The situation is slightly different if the group $G$ is a Lie group. Then again the quotient $Q = M/G$ does not need to be a manifold. To implement this for the propagation of the string, however, one is rather led to considering gauge theories, where in addition to the string coordinate $X$ one adds $\mathfrak{g}$-valued 1-forms $A$ on $\Sigma$, where $\mathfrak{g}$ is the Lie algebra of $G$. Let us consider a simple example, where $G=SO(2)$ acts in its defining representation on $M=\R^2$ as isometries when equipped with its standard flat metric $g= \dd x^2 + \dd y^2$. This is a somewhat degenerate example, but still sufficient to illustrate the principle. The quotient space of $M$ modulo the $G$-action, the orbits of which are circles centred around the origin, is a half-line, 
$Q := M/G \cong \R_{0,+}$. Thus it is a manifold with boundary at $0 \in \R_{0,+}$, which is the singularity of the quotient we are discussing in this example.\footnote{One also might add further real ``spectator'' coordinates $z_1, \ldots , z_{n-1}$; in this case the quotient space $Q$ would be an $n$-dimensional manifold with $(n-1)$-dimensional boundary. In more general cases such quotients need not be manifolds.} The metric $g = \dd r^2 + r^2 \dd \varphi^2$, in the adapted polar coordinates, then also induces a metric $g_N = \dd r^2$ on this half-line parametrized by $r \in  \R_{0,+}$.

Let our purpose be to in fact describe a propagation of our string in the quotient space $Q$. Since $Q$ is not a manifold, but a singular space---in the toy model here the singularity is very mild, it is only the boundary, but in general certainly it can be much worse---, we define a functional on $X \colon \Sigma \to M\equiv \R^2$, corresponding to two scalar fields $(X^i)_{i=1}^2:=(X,Y)\in C^{\infty}(\Sigma)$, extended by the 1-forms $A = A_\mu \dd \sigma^\mu$ as follows:
\begin{equation}\label{rot}
S_{gauged}[X,Y,A] := \frac{1}{2}\int_\Sigma \dD X \wedge \ast \dD X + \dD Y \wedge \ast \dD Y \, ,
\end{equation}
where we introduced the covariant derivatives 
\begin{equation}
\dD X \equiv \dd X + Y A~, \qquad \dD Y \equiv \dd Y - X A \, .
\end{equation}
{}Variation yields the equations $\dd * \dD X + A \wedge *\dD Y = 0$, $\dd * \dD Y - A\wedge *\dD X= 0$ together with the constraint $Y \dD X = X \dD Y$. Evidently, $X^2 + Y^2$ is a gauge-invariant quantity. The first two equations yield $\dd * \dd (X^2 + Y^2) = 2 \dD X \w *\dD X + 2 \dD Y \w *\dD Y$. Together with the constraint, a calculation then establishes 
\begin{equation} \label{Deltar}
\square \left( \sqrt{X^2 + Y^2} \right) = 0 \, ,
\end{equation} 
where $\square = \dd * \dd$ is the Laplacian or d'Alembert operator on the worldsheet $\Sigma$. This equation is the expected one for the propagation of the string in the radial direction. However, one has to note that in the above it was established only wherever $X^2+Y^2\neq 0$, since otherwise the non-derivability of the square root comes into play. 

On the other hand, the equations following upon variation from \eqref{rot} are perfectly well-defined for the string passing the origin $x=y=0$ in the target manifold $M$. To establish an equation like \eqref{Deltar} everywhere, including the singular origin, we may proceed as follows: Let us impose the gauge condition $Y=0$. Then $\dD X = \dd X$ and $\dD Y = - XA$ and the constraint simply turns into  $X^2 A = 0$, so that if $X$ has at most isolated zeros, the gauge field $A$ has to vanish. Thus the only non-vanishing field now is $X$ and its  field equation is simply 
\begin{equation}\label{Laplace}
\square X  = 0 \, .
\end{equation}
Noticing that the gauge condition $Y=0$ admits a residual gauge symmetry $X \sim - X$, we find the 
completely regularized version of Eq.~\eqref{Deltar}: Free movement along a real line, expressed by Eq.~\eqref{Laplace}, together with identification of positive and negative values of the field.

As simple as this example seems to be, it teaches us two more lessons. The first one is as follows: if the topology of $\Sigma$ is non-trivial, containing in particular non-contractible loops $l \subset \Sigma$, there can be additional gauge-invariant sectors that we did not see in the above consideration. Correspondingly, one may find a deviation from the naive picture of equivalence of the unreduced and the  reduced theory, and this even if the target space is \emph{not} singular. So, let us exclude the origin from the target manifold above, restricting to maps $X \colon \Sigma \to M_{reg} \equiv \R^2 / \{(0,0)\}$.\footnote{One may here either think of the full target manifold $\R^2$ in the sigma model, where such maps $X$ still exist, \emph{or} consider the target manifold $\R^2 / \{(0,0)\}$, then \emph{regularly} foliated by the concentric circles generated by the rotations to be gauged out.} Above we performed the gauge choice $Y(\sigma)=0$ for all $\sigma \in \Sigma$. In fact, this cannot be always attained: Consider a map $X \colon \Sigma \to M$ such that $X(l)$ has a non-trivial winding number around the origin in $\R^2$. Evidently this winding number cannot be changed by smooth deformations of $X$ along radial orbits in $M$ or $M_{reg}$---only for trivial winding numbers the gauge $Y \equiv 0$ can be attained.

Such winding numbers then show up for example in the parallel transport of $A$ around the non-contractible loop. 
To see this more explicitly, consider $\Sigma = S^1 \times \R$ and maps $X \colon \Sigma\to M_{reg}$ having a non-trivial winding number $k \in \Z^*$. Then we can change from Cartesian to polar coordinates $(r,\varphi)$ on the target. Denoting $R = X^* r$ and $\Phi = X^* \varphi$, the action then takes the form
\begin{equation}
S_{gauged} = \frac{1}{2}\int_\Sigma \dd R \wedge * \dd R + R^2 (\dd \Phi - A) \wedge * (\dd \Phi - A) \, .
\end{equation}
Variation with respect to $A$ yields $A = \dd \Phi$. Integrating this along $l$, we indeed obtain the gauge-invariant quantity $\int_l A = 2\pi k$. In fact, for non-trivial $k$, it is difficult to imagine that the string is moving on the half line only; or, to say it in another way, there would be different movements on the half line parametrized by an integer $k$, corresponding to different sectors of the theory. We will come back to this below.

It is remarkable that these additional sectors can be removed by realizing the half line $\R_0^+$ as a quotient of $\R^n/SO(n)\cong \R_0^+$ for any $n>2$, in otherwise the completely analogous way as the one for $n=2$ above. In general, it is to be expected that realizing a given singular space in different ways as a quotient of a smooth manifold by a group action or, more generally, by a foliation on it, can give different string theories in the end.

Second, returning to \eqref{rot}, we notice that  $A$ can carry also gauge invariant parameters independent of the map $X$. Consider maps $X \colon \Sigma \to M$ such that $X(\Sigma) = 0 \in \R^2$. Then the field equations resulting upon variation of \eqref{rot} become \emph{empty}. This means in particular, that $A$ is not at all constrained on $\Sigma$ and up to gauge invariance this sector of the theory yields the \emph{infinite dimensional} space of 1-forms on $\Sigma$ which are either co-exact or harmonic. This sector is in some sense unphysical: the string sits inside the singularity without moving. In this way, this part of the solutions can be well distinguished (and thus excluded if one prefers so) from the physically relevant ones, where the string really moves in the quotient space.

In the example that we now discussed in detail, the 2-form $B$ was set to zero. In general, this $B$-field can be thought of as an abelian 2-form gauge field describing a gerbe on $M$. Its curvature is a closed 3-form $H$, which globally is not necessarily exact. A typical example of this situation is described by the Wess-Zumino-Witten model \cite{WZW}, where the target is chosen to be a semi-simple group $G$ and the metric $g$ in \eqref{SgB} the bi-invariant extension of the Killing metric. For a particular choice of a matrix representation of $G$, the action is \begin{equation} \label{WZW}
S_{WZW}[g] = \frac{k\hbar}{8\pi} \left( \int_\Sigma \mathrm{tr}(g^{-1}\dd g \wedge \ast g^{-1}\dd g) +
\frac{2}{3} \int \mathrm{tr}(g^{-1}\dd g)^{\wedge 3}
\right) \, ,
\end{equation}
which for integer $k$ gives a single valued integrand in the exponent of the path integral.\footnote{We refer to \cite{WZW} for the discussion of Wess-Zumino terms. For the purpose of our paper it is moreover not necessary that the functional is single-valued, but only that the variational problem is meaningful. And indeed, the contribution to the variation of a WZ-term is always local.} This example is of interest also since it shows already potential obstructions in the gauging of its rigid symmetries \cite{WZW,Hull2,Figueroa,Figueroa2,Alekseev-Strobl}: for instance, it is not possible to gauge a subgroup $K \subset G$
acting by left or right multiplication, which then would yield the coset space $G/K$ as target. 
 
On the other hand, the adjoint action of any subgroup $K \subset G$ \emph{can} be gauged consistently, yielding the $G/K$-WZW models \cite{Gawedzki1,Gawedzki2}. Since the group action now has the identity element as a fixed point (similar to the origin in the simple model \eqref{rot} above) and does no longer act freely as does the left multiplication, the quotient space is also singular and such models describe strings propagating in the corresponding quotient space.  

For the same reason, the ``completely gauged'' $G/G$-WZW model still effectively has a non-trivial target space, while for this extreme case the model becomes topological upon the gauging. So, apparently in this last example, one somehow overdid the gauging and the string was implicitly frozen to no longer propagate in the directions that remain after taking the quotient. This is an important point to which we will come back below again.

\vspace{4mm}

In the present paper, we will go much further than just gauging the group action on a target manifold $M$ with respect to which the geometrical data, like $g$ and $B$ in \eqref{SgB}, are invariant. Instead of any action of something on $M$, we will consider the partition of $M$ into leaves forming a possibly singular foliation ${\cal F}$.
 Certainly group actions provide such a foliation, where the orbits are the leaves. In that case, however, \emph{any} leaf $L \subset {\cal F}$ has to be diffeomorphic to a coset space $G/K$ of the group $G$ acting on $M$ for some $K$, where $K$ is the stabiliser subgroup of a point in the leaf. This is evidently very restrictive since the different options for subgroups are not so abundant and does not become much better if one replaces the group action by the one of a Lie algebra action. (The situation changes, if one permits infinite dimensional groups certainly, but we will not consider this option here, at least not directly).

So, instead of on group actions, we just focus on the possibly singular foliation ${\cal F}$ chosen on $M$. We will then pose the question, what the geometrical data need to satisfy on $M$ such that one can define a gauge theory. The gauge theory is supposed to have the following minimal features, for a given foliation ${\cal F}$ on $M$: 
\begin{itemize}
\item The searched-for functional $S=S[X,A]$ depends on the string maps $X \colon \Sigma \to M$ and on additional 1-form gauge fields $A$. Putting the additional $A$-fields to zero, the action reduces to a functional $S[X,0]=S_0[X]$ which agrees with \eqref{SgB}, $S_0 \equiv S_{bos}$, or to a Wess-Zumino generalization of this.
\item The action $S$ has a gauge symmetry, which, on the scalar fields, reduces to arbitrary $\Sigma$-dependent deformations along the leaves of ${\cal F}$. (We will characterize this item more carefully below in terms of diffeomorphism groups). 
\end{itemize}
There is one third property that we might want to ask for, which is the following one:
\begin{itemize}
\item Whenever the foliation ${\cal F}$ is regular and the quotient $M/{\cal F}$ is a smooth manifold $Q$, we require that $Q$ can be equipped with a metric and a 2-form $g$ and $B$, respectively, such that for strings \emph{with a contractible topology of $\Sigma$} the dynamics following from $S$ modulo gauge invariance is the one of a variational problem \eqref{SgB} with target $Q$ (and likewise so for the case of a closed 3-form $H$ instead of $B$). 
This requirement may be extended to regular parts of any foliation ${\cal F}$, requiring the above in local regions which have good quotients then. 
\end{itemize}
Note that in the last item we required triviality of the world-sheet topology since without this condition, even in good cases like in the regularly foliated example with $M=\R^2/\{(0,0)\}$ acted upon by rotations, we would not have the required equivalence. There are also other occasions, where we do not encounter global obstructions, such as when replacing $\R^2/\{(0,0)\}$ modulo $SO(2)$ by $\R^n/\{(0,\ldots,0)\}$ modulo $SO(n)$ for some $n>3$. The reason there is that the homotopy classes of maps from a 2-surface into a sphere $S^n$ are always trivial then. So, if one wants, one may relax the condition on $\Sigma$ above for the cases that the leaves have trivial $\pi_1$ and $\pi_2$. On the other hand, one may require this property to only hold locally on $\Sigma$ for unrestricted $\Sigma$; this is the option that we will prefer henceforth.

For clarity we summarize the different options of gauging in the following way:
\begin{defn} \label{Def1}
A functional $S$ satisfying the  first two properties marked by a solid dot above is called a \emph{gauging of $S_0$ along the foliation ${\cal F}$}. 
\end{defn}
The third property marked by a dot is a bit more delicate since the \emph{same} functional can be used for different types of foliations. Let us introduce the following notation for singular foliations: Denote the maximal regular sub-foliation inside $M$ by ${\cal F}_{reg}$, which now foliates $M_{reg} \subset M$ regularly. For regular foliations certainly $M=M_{reg}$.  It may still be that there is no good quotient: think for example of an ordinary torus foliated by 1-dimensional leaves with irrational slope. So, for every non-regular foliation, the statement is to hold for all maps $X$ from (sufficiently small) discs $D \subset \Sigma$ mapping into small balls $B \subset M_{reg}$ that have a good quotient; the statements have to hold without any boundary conditions in this case. We now define:
\begin{defn} \label{Def2}
If the third property marked by a dot (interpreted as just explained) is satisfied in addition to the first two, we speak of a \emph{strict gauging} or a \emph{strict gauging outside singularities}.  Otherwise we speak of a \emph{non-strict} gauging.
\end{defn}

For example, the simple toy model \eqref{rot} above is a strict gauging (outside the singularity).\footnote{We clarified this only for one choice of the foliation, the one induced by rotations around the origin. For the full proof, performed in a more general context, we have to refer to section Sec.~\ref{sec:2.3} below.}
On the other hand, the $G/G$ WZW-model is a non-strict gauging of the WZW-model: While the completely gauged WZW model reduces to \eqref{WZW} when the gauge fields are put to zero (first condition above) and the gauge symmetries are parametrized by any $h \in C^\infty(\Sigma,G)$ such that any field $g \in C^\infty(\Sigma,G)$ is gauge equivalent to $h^{-1} g h$, as required for the adjoint orbits in $G$ (second condition above), the third requirement is certainly not fulfilled. Take, e.g., $G=SU(n)$. Its dimension is $n^2-1$ and its rank $n-1$. Thus the quotient space of $G$ modulo its adjoint action has dimension $n-1$; if one were to describe a propagation in this $n-1$ dimensional quotient space $Q$, the theory could not be topological---which, however, it is. 

We may illustrate this feature also at the toy model \eqref{rot} above. Suppose we introduce one more gauge field 1-form $\hat{A}$ and then add the following term to the functional \eqref{rot}:
\begin{equation}\label{rot'}
\widehat{S}_{gauged}[X,Y,A,\hat{A}] := S_{gauged}[X,Y,A] + \int_\Sigma \hat{A} \wedge \dd ( X^2 + Y^2) \, .
\end{equation}
Evidently \emph{both} of the first two conditions remain satisfied: If we set $A$ and $\hat{A}$ to zero, we get an action of the form \eqref{SgB}, namely for the choice $B=0$ and $g$ the standard metric on $M=\R^2$. Also the gauge symmetries are the old ones plus something that effects only the new gauge field, $\hat{A} \sim \hat{A} + \dd \epsilon$; so we keep the second property as well. However, variation with respect to $\hat{A}$ now freezes the radial movement of the string, and in contrast to \eqref{rot}, the model \eqref{rot'} is topological.\footnote{The gauge invariant content of \eqref{rot'} is not empty: even for trivial topology of $\Sigma$ one has the radius $r \in \R^+$ at which the string has to sit. For non-trivial topology of $\Sigma$, we get in addition holonomies of the gauge field $\hat{A}$ around non-trivial circles on $\Sigma$. But for sufficiently reasonable $\Sigma$, we obtain a \emph{finite-dimensional} solution space (modulo symmetries), which we will loosely call then a classically topological field theory. --- If we add spectator coordinates $z_1,\ldots, z_{n-1}$ to the target, then the theory \eqref{rot'} is no longer topological. Still, instead of effectively $n$ dimensions of the target as when added to \eqref{rot}, here one only has $n-1$ such directions left  (along which the string can propagate).} So, while \eqref{rot} is a locally strict gauging of $S_{bos}$ with target $\R^2$ and standard metric $g$, the gauge theory \eqref{rot'} is a non-strict gauging of it. 

Unwanted freezing of a movement transversal to those leaves which are factored out by gauging, and thus leading to a non-strict gauging, is one of the topics we will address in the present paper. On the other hand, additional winding numbers in a strict gauge theory, as discovered for the simple toy model \eqref{rot}, will not be discussed further in this paper. One may speculate, however, that it has in some sense an opposite effect than the freezing, leading to  ``emergent'' extra dimensions on the quantum level. Indeed, we are used from T-duality that winding modes can be turned into the discrete spectrum of a momentum for a movement of the string on a ``dual circle'' in the target. Since the original circles here are gauged out, such ``dual circles'' would correspond to new, compact directions of the string moving in a singular space-time. Such a phenomenon would occur precisely when the leaves of the foliation to be factored out are not simply connected. It would be interesting to explore this further elsewhere.

\begin{figure}[ht]
   \centering
\includegraphics[width=0.8\textwidth]{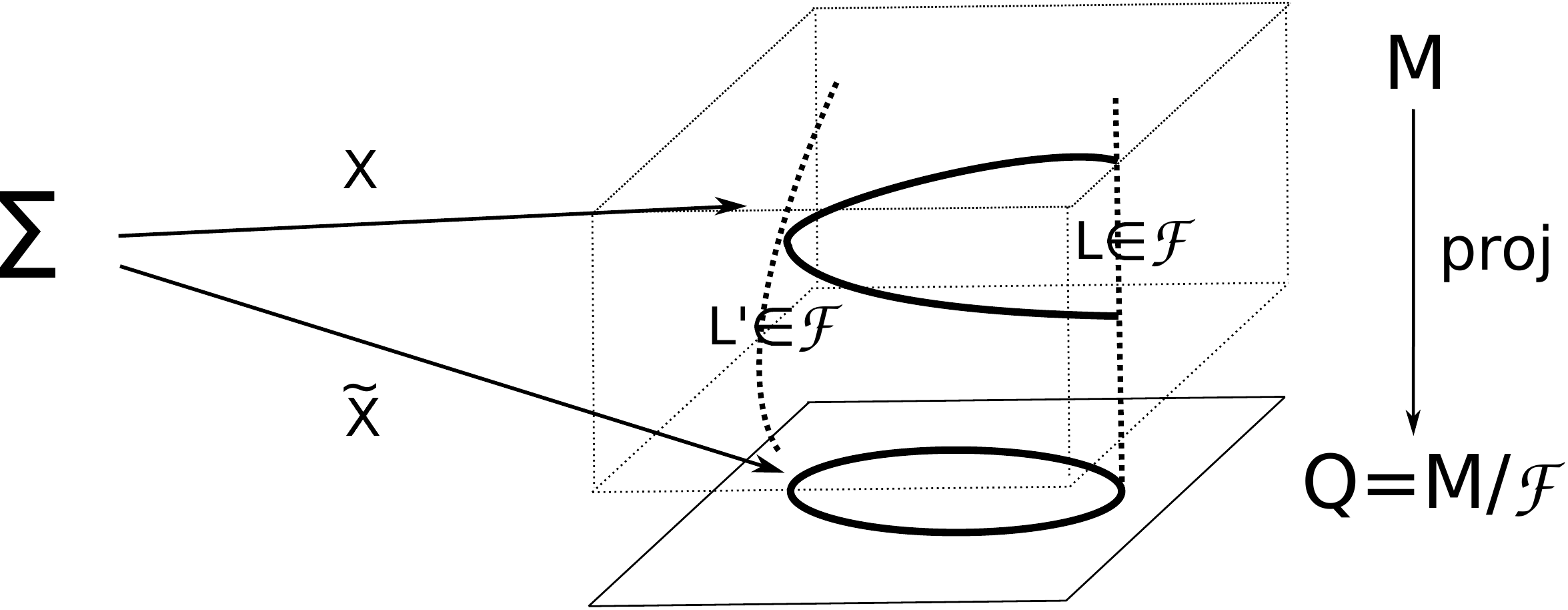}
   \caption{An open string in the mother space $M$ can turn into a closed one in the quotient $Q$; this happens precisely when its end points lie in the same leaf $L$ in $M$.}
    \label{fig:parallel}
\end{figure}

Another noteworthy feature of strings described by quotienting is the following one: An open string on $M$ can turn into a closed one after the quotient is taken. This happens precisely, when the two ends of the open string upstairs end on the same leaf $L \subset \cF$. This is illustrated in Fig.~\ref{fig:parallel} and also will not be discussed further in the present article.

There are some further technical requirements that we will pose on the foliation, so as to have sufficient control on it (for further details and a motivation on this issue, cf.~\cite{Stroblinprep,LG-Lavau-Strobl}). 
We assume that there exists the following sequence of vector bundles over $M$
\begin{equation}\label{FETM}
F \stackrel{t}{\longrightarrow} E \stackrel{\rho}{\longrightarrow} TM
\end{equation}
which is exact on the level of sections and such that the image of the map $\rho$ generates the foliation. In some cases, in particular if dropping the third condition in the above wish list, we may content ourselves with the anchored bundle $E$ over $M$, i.e.~drop the requirement that the bundle $F$ exists, which permits spanning the kernel of the anchor map $\rho$ in a controlled way.

In formulas, this implies the following. Denote by $(e_a)_{a=1}^r$ a local basis of $E$ and by $(b_I)_{I=1}^s$ a local basis of $F$. Then the vector fields $v_a := \rho(e_a)$ are tangent to the foliation ${\cal F}$ in $M$ everywhere and generate it. In particular, they are involutive, i.e.~locally we have structure functions $C^a_{bc}$ such that
\begin{equation}\label{C}
[v_a,v_b]=C^c_{ab}(x)\, v_c \, .
\end{equation}
In general, the $v_a$'s will be linearly dependent at a given point $x \in M$, reflecting the fact that the map $\rho$ can have a kernel. According to the required exactness, this kernel can be spanned by the image of sections in $F$. In other words, for any section $s \in \Gamma(E)$ such that $\rho(s) = 0$ there are locally functions $s^I$ on $M$ such that $s = s^I t(b_I)$. Since the anchor map does not need to have constant rank, the foliation ${\cal F}$ can be singular. It is, however, locally finitely generated as being covered by the bundle $E$, the existence of which we assume in any case.

For a group action, we can choose $E = M \times \mathfrak{g}$, where $\mathfrak{g}$ is the Lie algebra of the group acting on $M$ and $\rho$ corresponds to the map from $\mathfrak{g}$ to the vector fields on $M$ generating the action. One of the main messages of this paper is, however, to make us shift our perspective from group actions to foliations as the more fundamental notion in the context of gauge theories. It is not only a more general notion, but it in particular gives much more flexibility in the construction of gauge theories.

The gauge theory will depend on the string coordinates $(X^i)_{i=1}^n$, corresponding to the map $X \colon \Sigma \to M$, together with $r$ gauge field 1-forms $A^a$. Both together correspond to bundle maps $a \colon T\Sigma \to E$,  the base map of which reproduces the map $X$, the fiber map corresponds to the set of 1-forms (at least locally), where $r$ is the rank of $E$. Thus, according to the first requirement above, the searched-for gauged action $S=S[a]$ depends on this map $a$. According to the second requirement, the gauge symmetries infinitesimally take the form
\begin{equation} \label{deltaepsilon}
\delta_\epsilon X^i =  v^i_a(X)\epsilon^a \, ,
\end{equation}
lifting in one or the other way to $\delta_\epsilon a$ for all of $a$ such that $\delta_\epsilon S=0$. Here $\epsilon^a$ are the $\sigma$-dependent components of $\epsilon \in \Gamma(X^*E)$ and the above formula can be rewritten more abstractly as 
$\delta_\epsilon X = \rho(\epsilon)$. Note that $X$ enters this construction implicitly, which makes sense, since $\delta_\epsilon$ is to be thought of as a tangent vector on the (infinite dimensional) field \emph{space}  $\cA \equiv \{ a \}$ of bundle maps at the  given ``point'' $a$ at which the functional $S$ is evaluated. 

We introduce ``covariant derivatives'' by the usual formula, also in this more general context:
\begin{equation}\label{DX0} 
\dD X^i = \dd X^i - v_a^i(X) A^a~,
\end{equation}
or, more abstractly, $\dD X = \dd X - \rho(A) \in\Omega^1(\Sigma, X^*TM)$. Minimally coupled theories consist of theories where the gauge fields enter the theory only by means of the above covariant derivatives. In the context of WZ-terms, this is not sufficient, however; in fact, even when $B$ is not strictly invariant, but invariant only up to an exact term, minimal coupling does not provide a consistent gauge theory. 

In general the map $\rho \colon E \to TM$ has a non-trivial kernel. This is in particular a feature of singular foliations, where the rank of the anchor map $\rho$ can jump. Also there may be sections of $E$ which lie entirely in the kernel of this map, or one works for example over some open subset of $M$. Suppose for concreteness that $\psi \in \Gamma(E)$ is such a section such that $\rho(\psi) \equiv 0$. Denote by $A_\psi$ the corresponding 1-form gauge field. In the case of a minimally coupled
theory, the functional will not depend on $A_\psi$ at all. This means in particular that there is a simple shift symmetry of the functional $S[a]$ with respect to this field, $A_\psi \mapsto A_\psi + \lambda$ for any 1-form $\lambda$. This symmetry is not visible in the symmetries \eqref{deltaepsilon}, since only the ``$\rho$-shadow'' of $\epsilon$ enters the variation of $X$. 
Moreover, it \emph{cannot} be contained even in the lift of \eqref{deltaepsilon} to all of $a$ for simple form-degree reasons.

Let us make all this clearer, by taking care also of where precisely each field lives also. 
It is for example here where the usefulness of the map $t$ in the sequence \eqref{FETM} becomes transparent: Let $\lambda^I$ be the arbitrary, $\sigma$-dependent 1-form components of $\lambda \in \Gamma(T^*\Sigma \otimes X^*F)$. Then a minimally coupled theory has the feature that the action is invariant with respect to the transformations\footnote{It is noteworthy, that such a part of the symmetries appears also naturally in the context of higher gauge theories, cf.~\cite{Gruetzmann-Strobl} and \cite{Stroblinprep}, where in general the $\epsilon$ and $\lambda$ symmetries cannot be separated as easily as in the examples discussed directly hereafter.}
\begin{equation}\label{deltalambda}
\delta_\lambda A^a = t^a_I(X) \lambda^I~,
\end{equation}
i.e.~w.r.t.~$\delta_\lambda A = t(\lambda)$, where here $t$ denotes the map between the pullback bundles $X^*F$ and $X^*E$, induced by the map of the same name in \eqref{FETM}. The transformation \eqref{deltalambda} cannot be already contained in $\epsilon$ by lifting to $a$, since $\epsilon^a$ are functions on $\Sigma$ and $\lambda^I$ are 1-forms, as is needed for shifting the ``superfluous'' gauge fields. 

It is illustrative at this point to come back again to the simple toy model studied above. In the case of the action $S_{gauged}$, Eq.~\eqref{rot}, the bundle $E$ is a trivial line bundle over $M=\R^2 \ni (x,y)$, $E=M\times \R$ and the anchor $\rho$ is  evaluated at the unit section by means of the generator of rotations in the plane, $\rho(1) = x \partial_y - y \partial_x$. In this case, $F= M \times \{ 0 \}$ since the anchor map has a kernel only at the origin $(x,y)=(0,0)$, dropping in rank only at a single point. This cannot be resolved in terms of sections, since there is no non-vanishing section of a vector bundle vanishing outside of a single point. On the other hand, for the gauge theory $\widehat{S}_{gauged}$, Eq.~\eqref{rot'}, the bundle $E$ has rank $r=2$. Denote a basis of sections of $E$ by $\{ e, \hat{e} \}$ such that $\rho(e) = x \partial_y - y \partial_x$ and $\rho(\hat{e})=0$. The corresponding two 1-form gauge fields are $A$ and $\hat{A}$, respectively. 
Now the bundle $F$ has rank $s=1$. Let $F= M\times \R$ with unit section $b \cong 1$. 
 
In fact, we could already consider the action \eqref{rot} as depending on $A$ and $\hat{A}$, but, since minimally coupled, $\hat{A}$ does not appear in the functional $S_{gauged}$ and it has the $\lambda$-shift symmetry
\begin{equation}\label{lambdashift}
\delta_{\lambda} A = 0 \; , \quad \delta_{\lambda} \hat{A} = \lambda \, ,
\end{equation}
where we made use of the fact that $t(b)=\hat{e}$. 
The action $S_{gauged}$ is invariant with respect to 
the $\epsilon$-symmetry
\begin{equation}
\delta X = - Y \epsilon \; , \quad \delta Y = X \epsilon \; , \quad \delta_\epsilon A = \dd \epsilon
\end{equation}
with an \emph{arbitrary} prescription for $\delta_\epsilon \hat{A}$ (in terms of $\epsilon$ and $\hat{\epsilon}$) and, in addition, with respect to the $\lambda$-shift symmetry \eqref{lambdashift}. On the other hand, the action $\widehat{S}_{gauged}$ does no longer have this $\lambda$-symmetry; instead, the gauge field $\hat{A}$ is a second, independent $U(1)$ gauge field, transforming according to 
\begin{equation}
\delta_\epsilon \hat{A} = \dd \hat{\epsilon} \; .
\end{equation}
In the present example it is thus the $\lambda$-shift symmetry shared by $S_{gauged}$ but not by $\widehat{S}_{gauged}$ that ensures the third property we wish the gauge theory  to have so that the string propagates in the quotient target and does not become frozen in some directions. We will see that this simple observation in the toy model will hold also for the most general gauged model. 

In general, it is not so easy to disentangle ``necessary'' from ``unnecessary'' dimensions of (the fibers of) $E$. Consider for example rotations in $M=\R^3$ instead of in $\R^2$. The orbits are concentric 2-spheres around the origin, which is a 0-dimensional orbit by itself. If the origin is removed from the manifold, the foliation is regular (the leaves have constant dimension 2) and a minimal anchored bundle yielding this foliation of $M^*:=\R^3\backslash \{(0,0,0)\}$ has rank $r=2$ and is the subbundle of $TM^*$ of vectors parallel to the spheres. For all of $M$ and the singular foliation given by rotations, it is natural to consider the action Lie algebroid $E = M \times \mathrm{so(3)}$ so that the rank $r=3$. Let $e_i$ denote the basis vectors of $\mathrm{so(3)}$ or, better, of $\Gamma(E)$ generating a rotation around the $i$-th axis of $M=\R^3$. Then $\rho(e_i) = L_i \equiv \varepsilon_{ijk} x^j \partial_{k}$, the 
$i$-th component of the angular momentum operator (up to irrelevant prefactors). In this case, $F$ has rank $s=1$ and if $b$ denotes its unit section, then $t(b) = x^i e_i$. This then evidently fulfills $(\rho \circ t)(b) \equiv \rho(t(b))=0$ due to the antisymmetry of the $\epsilon$-tensor in three dimensions. We will come back to a similar situation when discussing the SU(2)-WZW model in the body of the paper. 

Before concluding this section, we come back to the second item in the desiderata of the gauging (cf.~Def.~\ref{Def1}). We already observed above that, in formulas and on the infinitesimal level, this requirement is expressed precisely by Eq.~\eqref{deltaepsilon}. We now intend to formulate it on the level of the infinite-dimensional gauge group $\cG$ leaving invariant the action functional $S$ gauging $S_0$. 

For this purpose we consider the trivial bundle ${\cal M} := \Sigma \times M \to \Sigma$. Denote by 
$\mathrm{Diff}_{vert}({\cal M})$ its vertical diffeomorphisms, i.e.~diffeomorphisms of the manifold ${\cal M}$ which do not move its fibers when considered as a bundle, or, in other words, which project to the identity map on its base $\Sigma$. By this trick (cf.~also \cite{BKS}) we obtain a mathematical description of diffeomorphisms of $M$ parametrized smoothly by the worldsheet manifold $\Sigma$. What we are interested in is a subgroup of these diffeomorphisms leaving invariant the foliation $\cF$ of $M$. We denote this group as follows
\begin{equation}
\cG_\cF := \mathrm{Diff}^\cF_{vert}({\cal M})
\end{equation}
Any vector bundle morphism $a \in \cA \equiv {a \colon T\Sigma \to E}$ projects to a base map $X \in \cX:=\{X \colon \Sigma \to M\}$ and the functional $S$, the gauging of $S_0$, is a (local) functional on $\cA$. $\cG$ is supposed to act on $\cA$ and to leave $S$ invariant. Since $\cA$ is a space of vector bundle morphisms, the operation should also project to the base, and thus $\cG$ acts also on $\cX$. Some of the elements in $\cG$ will act on non-trivially on the fibers of $E$ and thus its action on $\cX$ is then trivial. What we require now is that $\cG$ is an extension of $\cG_{\cF,0}$, the identity component of $\cG_{\cF}$
\begin{equation}\label{GF}
\cG \stackrel{\mathrm{pr}}{\twoheadrightarrow} \cG_{\cF,0},
\end{equation}
such that for any $g \in \cG$ and $a \in \cA$ with basemap $X\in \cX$ one has
\begin{equation}
g \cdot a = \mathrm{pr}(g) \cdot X \,. 
\end{equation}
The projection $\mathrm{pr}$ is surjective, such that \emph{all} elements in $\cG_{\cF,0}$ are realised as gauge transformations on $\cX$. We restricted to the identity component here, since this is what can be reached by integrating infinitesimal transformations such as \eqref{deltaepsilon}.

In general, the map $\mathrm{pr}$ will have a huge kernel certainly. On the infinitesimal level this corresponds to elements $\epsilon \in \Gamma(X^*E)$ with $\rho(\epsilon) = 0$ as well as the $\lambda$-transformations we discussed above. We remark, however, that these two parts do not need to be independent from one another. To give a simple example, consider the action functional $S\equiv0$ viewed as a functional on $a \colon T\Sigma \to \R$, where $\R$ is the vector bundle $E$ over a point. The vector bundle morphisms $a$ correspond then simply to 1-forms $A \in \Omega^1(\Sigma)$ in this case. $\R$ is an abelian Lie algebra and the conventional symmetries on this $A$ are of the form $A \mapsto A + \dd f$, for an arbitrary $f \in \C^\infty(\Sigma)$. On the other hand, the $\lambda$-symmetry consist of arbitrary shifts of $A$ in this case, $A \mapsto A + \lambda$ for any $\lambda \in \Omega^1(\Sigma)$. So, in this example, the (non-acting) $\epsilon$-symmetries form even a subgroup of the $\lambda$-translations. In this simple example, one has $\cG \cong (\Omega^1(\Sigma),+)$, while $\cG_\cF = \{ 1 \}$ is the trivial group.

Let us come back to the structure of the remaining part of this article: In Section \ref{Sec2} we consider only theories that are minimally coupled. As such they will have full $\lambda$-shift symmetry whenever the bundle $E$ is chosen ``unnecessarily big'' and there is no freezing of the string propagation. The construction is, however, not limited to group actions by any means; instead we will determine the most general conditions on $g$ and $B$ such that a foliation fixed by means of the anchored bundle $E$ over $M$ together with anchor $\rho \colon E \to TM$ can be gauged (exhibiting all the three above-mentioned features required for a gauging).
Already in this context, we will find a mixing of the conditions on $g$ and $B$, in general, a feature restricted to two-dimensional sigma models since only there the Hodge-dual of a 1-form is a 1-form again.

In Section \ref{Sec3} we then turn to a more general scenario, like the one where the $B$-field corresponds to a non-trivial gerbe with closed 3-form curvature $H$. But even for ordinary group actions on $M$ and in the presence of merely a $B$-field that is not strictly invariant with respect to the $\mathfrak{g}$-action but changes by the exterior derivative of a 1-form on $M$, minimal coupling is not sufficient and the question of the form and the properties of a gauged action functional needs to be addressed. We will do this in a pedagogical way in Section \ref{Sec:3.1}, paving the way for the general discussion in what follows. In Section \ref{Sec:3.2} we then address the gauging of the standard 2d sigma model with metric $g$ twisted by a closed 3-form $H$: We restrict the kinetic sector to minimal coupling, containing all the $\gamma$-contributions to the gauged functional, but make an otherwise most general ansatz for the gauging of the Wess-Zumino term. Under this assumption, we will find the most general conditions on $g$ and $H$ such that the theory can be gauged, and provide the corresponding form of the gauged action and its gauge symmetries. (Dropping the restriction of minimal coupling in the kinetic sector is deferred to Appendix \ref{AppendixA}, since it is more technical).

In the final subsection, Sec.~\ref{Sec:3.4}, we reformulate the findings of Section \ref{Sec:3.2} in terms of generalized geometry and Dirac structures. As mentioned already in the introduction, cf.~in diagram~\eqref{diagrammain}, we show in particular, that for any data fixing an ungauged model there is a universal functional $S_{univ}$ such that whatever the chosen admissible singular foliation $\cal F$ on $M$ is, the correspondingly gauged action $S=S_{gauged}$ follows from a pull-back:  $S_{gauged} = \widehat{\sigma}^* S_{univ}$. 



\section{Minimal Coupling and Mixing of g and B for Strings} \label{Sec2}

In this section we discuss the most general conditions on the metric $g$ and the 2-form $B$ for a given singular foliation such that minimal coupling provides a gauge theory in the sense specified in the previous section. While for sigma models of a dimension $\dim \Sigma \equiv d > 2$, both $g$ and $B$ have to satisfy two  equations independent from one another, coupling only indirectly due to the use of the same connection $\nabla$ in $E$ used to express the generalized invariance, for strings, $d=2$, the conditions of $g$ and $B$ will be found to mix directly, as typical for the appearance of generalized geometry, an appearance that will become even more pronounced in the subsequent section.
Before showing this in detail, however, we will briefly recall the conventional setting for gauging.

\subsection{Conventional gauging of group actions}
\label{Sec:2.1}
The usual procedure of gauging comprises a rigid (global) symmetry, which is made local by means
of minimal coupling. 
For example, let us consider the ungauged action
\be \label{S01}
S_0[X]=\int_{\S}\sfrac 12 g_{ij}(X)\dd X^i\w\ast \dd X^j~.
\ee
In the presence of an isometry for the metric, this action can be gauged with well-known methods. 
Indeed, consider the Lie algebra $\mathfrak{g}$ with elements $\xi$ and assume that 
\be \label{Lie0}
{\cal L}_{v}g=0~,
\ee
for  vector fields $v=\rho(\xi)$,  $v=v^i\partial_i$.
Then consider Lie algebra valued 1-forms $A=A^a\xi_a$ and the minimal coupling 
\eqref{DX0} 
where here $v_a \equiv v_a^i \partial_i= \rho(\xi_a)$ are the fundamental vector fields on $M$ corresponding to a basis $\xi_a$ of the Lie algebra.
The corresponding gauged action reads
\be \label{standardgaugedaction}
S[X,A]=\int_{\S}\sfrac 12g_{ij} \dD X^i\w\ast \dD X^j~.
\ee
We already considered a special case of this construction in the Introduction; there $(M,g)$ was $\R^2$ equipped with its standard metric, and we gauged the rotations, i.e.~$\mathfrak{g}=\R$ and $v\equiv\rho(1)= x \partial_y - y \partial_x$.
 The action (\ref{standardgaugedaction}) is invariant under the infinitesimal gauge transformations
\bea \label{gauge1a}
\d_{\epsilon}X^i&=&v^i_a(X)\epsilon^a~,\\
\label{gauge1b}
\d_{\epsilon}A^a&=&\dd \epsilon^a+C^a_{bc}A^b\epsilon^c~,
\eea
where $\epsilon$ is a gauge transformation parameter, depending arbitrarily on $\Sigma$, and $C^a_{bc}$ are the structure constants of the Lie algebra $\mf{g}$.
This is easily proven as follows. First recall that the Lie derivative is given in components as
\be \label{Lieg}
({\cal L}_{v}g)_{ij}=v^k\partial_kg_{ij}+g_{kj}\partial_iv^k+g_{ki}\partial_jv^k~.
\ee
Direct variation of the action yields
\bea 
\d_{\epsilon}S&=&
\int_{\S}\biggl(\sfrac 12({\cal L}_{{v}_{a}}g)_{ij}\epsilon^a\dD X^i\w\ast \dD X^j+ g_{kj}\epsilon^a\big([v_b,v_a]_{\text{Lie}}^k-
C_{ba}^cv_c^k\big)A^b\w\ast \dD X^j\biggl)~,
\eea
which means that 
\be
\d_{\epsilon}S=0\quad \Longleftrightarrow \quad {\cal L}_{v}g=0\quad \text{and} \quad [v_a,v_b]_{\text{Lie}}=C_{ab}^cv_c~,
\ee
as required. Note that we did not need the fact that $C^a_{bc}$ satisfy a Jacobi identity; this will play a role also later when using an almost Lie algebroid structure compatible with the anchored bundle $E$ in the context of gauging foliations.

This scenario generalizes in a straightforward way to the string moving in a $B$-field background as in \eqref{SgB}, i.e.~with the ungauged action 
\be \label{S02}
S_0[X]=\frac 12\int_{\S}g_{ij}(X)\dd X^i\w\ast \dd X^j+B_{ij}(X)\dd X^{i}\w\dd X^{j}~.
\ee
If in addition to \eqref{Lie0}, also $B$ is strictly invariant with respect to the group action, ${\cal L}_v B = 0$, the action functional \eqref{S02} has a rigid invariance again that can be gauged by minimal coupling:
\be \label{S12}
S[a] \equiv S_1[X,A]=\frac 12 \int_{\S}g_{ij}(X)\dD X^i\w\ast \dD X^j+B_{ij}(X)\dD X^{i}\w\dD X^{j}~,
\ee
where again $\dD X$ denotes the covariant derivatives \eqref{DX0} and $a \colon T\Sigma \to M \times \mathfrak{g}$ corresponds to the pair $(X,A) \in C^\infty(\Sigma,M)\times \Omega^1(\Sigma,\mathfrak{g})$. With a similar calculation as the one above, one shows that also the extended gauged functional \eqref{S12} is invariant with respect to the infinitesimal local (or gauge) symmetries \eqref{gauge1a}, \eqref{gauge1b}: $\delta_\epsilon S=0$ for any choice of $\epsilon^a \in C^\infty(\Sigma)$. 

Let us briefly compare this to the setting of \eqref{GF}. In the present context the \emph{structure group} $G$ corresponds to the integration of the Lie algebra $\mathfrak{g}$ (we assume here that the Lie algebra action gives rise to a group action). Then the conventional \emph{gauge group} takes the form $\cG_{gauge} \equiv C^\infty(\Sigma, G)$ here. As the discussion after \eqref{GF} shows, the invariance group of $S$ can be still bigger than this and in general one only knows $\cG_{gauge} \subset \cG$. Restricting the map $\mathrm{pr}$ to this subgroup, we obtain a morphism into $\cG_cF$, which in general has a kernel; on the level of Lie algebras, this kernel corresponds to those elements $\epsilon \in C^\infty(\Sigma,\mathfrak{g})$ that are annihilated  by $\rho$ (extended to the mapping space from $\rho \colon \mathfrak{g} \to \Gamma(TM)$).


In fact, at least if $\S$ has no boundary, the action  \eqref{S02} is also invariant with respect to the rigid 
transformations $\delta_\epsilon X^i = v_a^i(X) \epsilon^a$ for \emph{constant} $\epsilon^a$,  if 
${\cal L}_{v_a} B = \dd \beta_a$ for some $1$-forms $\beta_a$ on $M$. However, in this case, the functional \eqref{S12} 
then \emph{no longer} provides a gauge-invariant extension of \eqref{S02}: $\delta_\epsilon S \neq 0$ (even for 
constant $\epsilon^a$s) since the terms containing $\beta_a$ then no longer assemble into an exact contribution to the functional. 
We will come back to this issue in section \ref{Sec3} below.

\subsection{Minimal gauge theory quotienting a  background with $g$ and $B$ along a singular foliation} \label{Sec:2.2}

In this section the ungauged and the gauged theory take precisely the same form  as in the previous subsection, Eqs.~\eqref{S02} and \eqref{S12}, respectively, where now, in generalization of the above, $a \colon T\Sigma \to E$ and $X \in C^\infty(\Sigma,M), A \in  \Omega^1(\Sigma,X^*E)$. The main difference is that we do no longer require a group action on $M$, but instead we are given a singular foliation generated by the involutive vector fields $v_a$ entering the action in terms of the covariant derivatives \eqref{DX0}.  While certainly the group orbits provide a singular foliation, by no means every foliation results from a group action.

As explained in Sec.~\ref{Orientation}, we assume this foliation to fit into the sequence \eqref{FETM}, such that $\Gamma(F) \stackrel{t}{\longrightarrow} \Gamma(E) \stackrel{\rho}{\longrightarrow} \Gamma(TM)$ is exact. To make sense of formulas such as \eqref{gauge1b} in the absence of a Lie bracket as before, we will furthermore equip $E$ with an almost Lie algebroid structure, which means by definition that we assume that there is an antisymmetric bracket $[\cdot, \cdot]$ on the sections $\Gamma(E)$ such that $[s,fs']=f[s,s'] + \rho(s)f \, s'$ and such that $\rho([s,s']) = [\rho(s),\rho(s')]$. The latter equation ensures that in a local basis of sections $(e_a)_{a=1}^r$ with $\rho(e_a)=v_a$, one has 
\begin{equation} \label{structurefunctions}
[e_a,e_b] = C^c_{ab}(x) e_c \, ,
\end{equation}
with a choice of the structure functions compatible with \eqref{C}. Note that equation \eqref{C} does \emph{not} yet fix these structure functions, since the $v_a$s are linearly dependent in general. It is comforting to know that \emph{every} anchored vector bundle
$(E,\rho)$ over $M$ \emph{can} be equipped with such an almost Lie algebroid structure, cf.~\cite{LG-Lavau-Strobl}. 
To avoid any confusion, we remark also that the bracket $[ \cdot , \cdot ]$ is not required to satisfy a Jacobi 
identity---for which reason the algebroid $(E,\rho,[\cdot,\cdot])$ is called \emph{almost Lie} only---while still the image of the Jacobiator with respect to $\rho$ has to vanish, since the anchor map was assumed to be a morphism of brackets and the bracket on the vector fields is a Lie bracket certainly.

We now ask ourselves under which conditions does the functional \eqref{S12} have a gauge symmetry of the base map $X \colon \Sigma \to M$ along the given singular foliation, i.e.~infinitesimally of the form \eqref{deltaepsilon}. We do not want to make \emph{any} assumptions about the gauge transformations of $A$ at this point, we only want that the transformation $\delta_\epsilon X$ given by the formula \eqref{deltaepsilon} can be (non-uniquely) lifted to \emph{some} infinitesimal variation $\delta_\epsilon a$ for the map
$a$, such that $\delta_\epsilon S_1[a] = 0$ for all $\epsilon^a \in C^\infty(\Sigma)$.\footnote{It is clear at this point how to rewrite the ``for all''-part of this sentence in a more covariant form: $\epsilon \equiv \epsilon^a e_a \in \Gamma(X^*E)$ where $e_a$ denotes a local frame in $X^*E$ here.} We thus make the general ansatz 
\begin{equation}\label{deltaADelta}
\delta_{\epsilon} A^a=\dd \epsilon^a+C^a_{bc}(X)A^b\epsilon^c + \Delta A^a
\end{equation}
for a yet undetermined part $\Delta A^a$, which is only required to be linear in $\epsilon$ and its derivatives at this point. 

So the question posed in this subsection then has two parts: First, and most importantly, what are the conditions on $g$ and $B$ such that such a lift of symmetries exists? And second, how do the 1-form gauge fields transform in this case?

To tackle these questions, we first start with the transformation property of the covariant derivatives \eqref{DX0}. Making use of only the involutivity equation \eqref{C} at this point, they transform as follows:
\begin{equation} \label{deltaDX}
\delta_\epsilon \dD X^i = \epsilon^a v_{a,j}^i \dD X^j - v_a^i(X) \Delta A^a \, .
\end{equation}
 Note in particular that the contribution where the de Rham differential acts on $\epsilon$ cancelled out due to the parametrization 
\eqref{deltaADelta}, while in principle there can be still a $\dd \epsilon$-dependence of  $\Delta A^a$ at this point. Comparing this formula with the variation of $\dd X^i$ under the condition that $\epsilon^a$ are constants, we see that the first part of \eqref{deltaDX} is precisely of this form. Thus the variation of the two parts of the action \eqref{S12} containing $g$ and $B$ recombine into Lie derivatives as before, in addition to the terms containing the new contribution $\Delta A^a$ from \eqref{deltaDX}:\footnote{If one does not buy this argument, one may certainly just use the explicit formula \eqref{Lieg} and a similar one for the Lie derivative of $B$ to arrive at the result below, cf.~also \cite{DSMfourofus}. For notational compactness we dropped writing the dependence on $X$ of these Lie derivatives explicitly.}
\begin{eqnarray}
\delta_\epsilon S[a]&=& \int_{\S}\epsilon^a\biggl(  \sfrac 12({\cal L}_{v_a}g)_{ij}\dD X^i\w\ast \dD X^j+\sfrac 12 ({\cal L}_{v_a}B)_{ij}\dD X^{i}\w  \dD X^{j}\biggl) \nonumber \\
&&\!\!\!\!\!\!\!\!\! - \int_{\S} g_{ij} v_a^i \Delta A^a \wedge  \ast \dD X^j + B_{ij}v_a^{i} \Delta A^a \wedge \dD X^{j}\, . \label{deltaS}
\end{eqnarray}
We note en passant that for gauge invariance along the foliation one has
\begin{equation}
\Delta A^a = 0\qquad \Leftrightarrow \qquad {\cal L}_{v_a}g = 0 \quad \mathrm{and} \quad {\cal L}_{v_a}B= 0  \, . 
\end{equation}
In other words, without new contributions to the gauge transformation of the $A$-fields, we necessarily need strict invariance of $g$ and $B$ along a set of vector fields $v_a$ generating the foliation. While for a $2$-form $B$ such a condition is not extremely restrictive, for a metric $g$ it is; it implies in particular that there is a \emph{finite dimensional} Lie algebra $\mathfrak{g}$ acting by means of fundamental vector fields $v_a$ on $M$ and, simultaneously,  that the metric $g$ is one of the exceptional choices of it that have a non-trivial local isometry group, the Lie algebra of which contains $\mathfrak{g}$. In this case, we are back to the  conventional situation of gauge theories with a symmetry of the original action functional \eqref{S02} and a gauging by Lie algebra valued 1-forms $A$. 

Let us now consider the most general choices in \eqref{deltaS} ensuring $\delta_\epsilon S = 0$.  While for $d>2$,  evidently the contributions proportional to $g$ and its derivatives cannot cancel those of $B$ and its derivatives and vice versa, precisely for two-dimensional world-sheets $\Sigma$ describing strings, the Hodge dual of a 1-form is again a 1-form, and thus we have the option
of refining our ansatz 
\eqref{deltaADelta} by 
\be \label{Delta} \Delta A^a =  \omega^a_{bi}(X) \epsilon^b\dD X^i + \phi^a_{bi}(X) \epsilon^b\ast \!\dD X^i+\Delta_t A^a~,
\ee for some yet undetermined coefficients $\omega^a_{bi}(x)$ and $\phi^a_{bi}(x)$ on a patch of $M$ over which $e_a$ is a frame in $E$. Note that without any further information about $\Delta_t A^a$, this does not pose \emph{any} restriction on $\delta A$ at this point.

We now find that the variation \eqref{deltaS} vanishes \emph{if and only if}\footnote{$\wedge$ and $\vee$ denote the wedge product and its symmetric counterpart. Conventions are such that for example for any $\alpha$, $\beta$ $\in \Omega^1(M)$ one has $\alpha \wedge \beta = \alpha \otimes \beta - \beta \otimes \alpha$ and $\alpha \vee \beta = \alpha \otimes \beta + \beta \otimes \alpha$.}
the following conditions on $g$ and $B$ hold true:
\bea 
\label{condition1gB} {\cal L}_{{v}_a}g&=&\o^b_a\vee\iota_{{v}_b}g
-\phi^b_a\vee \iota_{{v}_b}B~,
\\
\label{condition2gB} {\cal L}_{{v}_a}B&=&\o^b_a\w\iota_{{v}_b}B\pm \phi^b_a\w \iota_{{v}_b}g~
\eea 
and, simultaneoulsy, the transformation on $A$ is constrained by
\begin{equation}\label{Deltaconstraint}
v_a^i(X) \Delta_t A^a = 0 \quad , \qquad \forall i=1,\ldots,n \, .
\end{equation}
Here we used the fact that $\ast^2=\mp 1$, depending on the signature of spacetime: The upper sign here as well as in \eqref{condition2gB} applies when the metric on $\Sigma$ is Euclidean, while for Lorentzian signature it is the respective lower one.

Every contribution $\Delta_t A$ satisfying the constraint \eqref{Deltaconstraint} remains completely undetermined by the requirements of invariance of $S$, since it drops out from the variation \eqref{deltaS}. On the other hand, by means of our assumptions on \eqref{FETM}, on any part of $M$ where $\rho$ has a constant rank, elements in the kernel of $\rho$ can be generated by the image of $t$. We thus may write
\begin{equation}
\Delta_t A^a = t^I_a(X) \lambda_I + \Delta_{sing} A^a
\end{equation}
for arbitrary 1-forms $\lambda_I$ on $\Sigma$, which may in particular also depend on $\epsilon$ and its derivatives. One might believe that with our assumptions on the exactness of the sequence \eqref{FETM} \emph{on the level of sections} the contributions $\Delta_{sing} A$ must necessarily vanish. However, $A$ is a 1-form valued section in the pullback bundle $X^* E$ only, not in $E$ itself. 
Let us now denote by $M_{sing}$ the subset of points $x \in M$ not admitting any neighborhood $U_x$ in which the rank of $\rho$ is constant. For any map $X \colon \Sigma \to M_{sing} \subset M$ such that $\mathrm{ker} \rho \subset E|_{M_{sing}}$ is a subbundle, there are elements $\alpha \in X^* \mathrm{ker} \rho$ for which $\alpha \not \in X^* t(F)$.

It is not completely clear to us, if such contributions to the symmetries are admissible in this more general context of a gauge theory. This requires also questions of analysis on the space of fields, which we will not enter here. Let us stress, however, that $M_{sing}$ is not to be confused with the generically singular quotient space $Q = M / \!\sim$ in which the string propagates effectively. It rather corresponds to singular points or regions in $Q$ in which the string can get stuck. Reinterpreting the toy model \eqref{rot} as such a gauge theory with target $\R^2 \times \R$, $M_{sing} = (0,0)\in \R^2$ and potential contributions $\Delta_{sing} A$ would permit to reduce the ``size'' of the \emph{space of solutions} having $X(\Sigma) = (0,0)$, strings stuck in $0 \in \R_0^+ \cong N$, to a single point. (We recall that this space was blown up in the standard treatment of the theory \eqref{rot} discussed in Sec.~\ref{Orientation} since $A$ is not restricted at all for a map $X$ that vanishes identically).

Excluding henceforth eventual ``singular'' 
(``skyscraper-type'') contributions to the gauge transformations, we thus find that the most general gauge transformations yielding the invariance conditions \eqref{condition1gB} and \eqref{condition2gB} have the form
\begin{eqnarray}
\label{deltaXA} \delta X^i \equiv \d_{(\epsilon,\lambda)}X^i&=&v^i_a(X)\epsilon^a~,\\
\d A^a \equiv \d_{(\epsilon,\lambda)} A^a&=&\dd \epsilon^a+C^a_{bc}(X)A^b\epsilon^c+ \omega^a_{bi}(X)\epsilon^b \, \dD X^i  + \phi^a_{bi}(X) \epsilon^b\ast\! \dD X^i+ t^I_a(X) \lambda_I~, \nonumber
\end{eqnarray}
for $\epsilon \in \Gamma(X^*E)$ and $\lambda \in \Gamma(X^*F)$ arbitrary and for some as yet undetermined coefficient functions $\omega^a_{bi}$ and $\phi^a_{bi}$. Note that any $\epsilon$-contribution to $\lambda$ can be absorbed by the unrestricted choice of $\lambda$. 

The gauge symmetries \eqref{deltaXA} will also play a role in Sec.~\ref{Sec3} below. There we will not always be able to require invariance with respect to all of $\lambda$. We thus introduce the simplified notation $\d_\epsilon$ for $\d_{(\epsilon,0)}$ and, similarly, $\d_\lambda \equiv \d_{(0,\lambda)}$ and call $\epsilon$-invariance simply gauge invariance and invariance with respect to $\delta_\lambda$ a $\lambda$-invariance or invariance with respect to $\lambda$-translations.

Each of the conditions \eqref{condition1gB} and  \eqref{condition2gB} contains both $g$ and $B$ on its r.h.s. This is typical for two dimensions and also for generalized geometry. We will come back to this perspective below as far as it concerns the construction of gauge theories, but for the purely geometrical interpretation of these formulas we refer to \cite{Killing2}. 
In particular, in the latter reference, the conditions are viewed as a single condition on a ``generalized metric'' (cf. \cite{Hitchin03,Kotov-Strobl10}). What is important to note here already, however, is that these conditions are \emph{much} weaker than (strict)  invariance of $g$ and $B$ with respect to a group action.

Before closing this subsection, however, we still want to rewrite the above two equations \eqref{condition1gB} and \eqref{condition2gB}
in a frame-independent manner. For this purpose we need to understand the global nature of the coefficients $\omega_a^b \equiv \omega_{ai}^b(x) \dd x^i$ and $\phi_a^b\equiv \phi_{ai}^b(x) \dd x^i$. While the two invariance conditions restrict the behavior of the coefficients with respect to changes of the frame, they do not determine it completely. However, the transformations \eqref{deltaXA} do: 
Consider a change of coframe $e^a \mapsto \widetilde{e^a} \equiv M^a_b(x) e^b$ corresponding to a change of $A^a$ to $\widetilde{A^a} \equiv M^a_b(X) A^b$. We require that in the new frame the gauge transformations for 
$\widetilde{A^a}$ again take the form of \eqref{deltaXA}, with all quantities replaced by the quantities with tilde on the r.h.s., corresponding to the new frame.\footnote{In particular, this concerns the unorthodox change of the structure functions $C^a_{bc}$ following from \eqref{structurefunctions} 
with respect to a change of the frame---defined in compatibility with \eqref{C}, which already permits to deduce this behavior in the case where the $v_a$s are linearly independent.} On the other hand, we have 
\begin{equation}
\delta_\epsilon \left( M^a_b(X) A^b \right) =  M^a_b(X)  \delta_\epsilon A^b +  \left(v_c( M^a_b)\right)(X) \:\epsilon^cA^b \: . \label{deltaMA}
\end{equation}
 Following \cite{BKS}, we are able to conclude from a direct comparison that $\omega^b_a$ are the coefficient 1-forms of a connection $\nabla$ in $E$, 
\begin{equation}
\nabla e_a = \omega_a^b \otimes e_b \, .
\end{equation} 
This then taking care of the inhomogeneity in the r.h.s.~of equation \eqref{deltaMA}, we find that $\phi_a^b$ transforms homogenously, i.e.~that 
its global meaning is a section $\phi \in \Gamma(T^*M \otimes \mathrm{End}(E))$, such that locally $\phi = \phi_{ai}^b(x) \,\dd x^i \otimes e^a \otimes e_b$. 
 
Now we are in the position of reformulating  Eqs.~\eqref{condition1gB} and \eqref{condition2gB} in an index-free way. Denote by $\rho\equiv v_a^i e^a \otimes \partial_i$ the section in $\Gamma(E^* \otimes TM)$ corresponding to the anchor map denoted by the same letter, $\rho \colon E \to TM$, and by $\nabla$ the connection on tensor powers of $E$ and $TM$ that is induced by $\nabla$ on $E$ and the Levi-Civita connection of $g$ on $TM$. Let $\phi(\rho)=\phi^b_a \otimes e^a \otimes v_b \in \Gamma(T^*M \otimes E^* \otimes TM)$ and $\iota_{\cdot} \cdot$ denote the contraction of the $TM$-part of a section put into the place of the first dot with the first $T^*M$-slot of the section following in the place of the second dot. So, for example, $\iota_\rho g = e^a \iota_{v_a} g$. Then the two extended invariance conditions \eqref{condition1gB} and \eqref{condition2gB} on $g$ and $B$ can be rewritten as 
\bea 
\label{condition1'gB} \mathrm{Sym} \left( (\iota_\rho \circ \nabla + \nabla \circ \iota_\rho ) g + \iota_{\phi(\rho)} B \right) &=& 0~,
\\
\label{condition2'gB} 
 \mathrm{Alt} \left( (\iota_\rho \circ \nabla + \nabla \circ \iota_\rho ) B \mp \iota_{\phi(\rho)} g \right) &=& 0~.
\eea 
Certainly, in the first equation we can drop the first appearance of $\nabla$ since $g$ is covariantly constant with respect to the Levi-Civita connection. This leads to the alternative form 
\begin{equation} \label{Sym}
\mathrm{Sym} \nabla \bar{\rho} = - \mathrm{Sym} \: \iota_{\phi(\rho)} B \, ,
\end{equation}
where $\bar{\rho} \equiv \iota_\rho g \equiv e^a \otimes g(v_a, \cdot)\in \Gamma(E^* \otimes T^*M)$, which resembles much more the usual Killing equation---to which it indeed reduces when $\phi =0$, $\nabla$ is flat, and $e_a$ is a covariantly constant frame. Likewise, the second equation can be rewritten by means of a ``covariantized Cartan formula'' using the exterior covariant derivative $\dD$ on $\Omega^\bullet(M,E^*)$ induced by the connection $\nabla$ on $E$ (and without any Levi-Civita part, which anyway drops out from \eqref{condition2'gB} by the anti-symmetrisation due to torsion-freeness) as
\begin{equation} \label{Alt}
(\iota_\rho \dD + \dD \iota_\rho ) B = \pm \mathrm{Alt} \: \iota_{\phi(\rho)} g \, .
\end{equation} 
Also this equation reduces to the ordinary invariance condition ${\cal L}_{v_a} B=0$ in the case of $\phi=0$, $\nabla$ flat and $e^a$ constant. 
While the equations \eqref{Sym} and \eqref{Alt} are attractive in their own right and adapted to the extension of a Killing symmetry of $g$ and an invariance of $B$, respectively, the form of the couple of Equations \eqref{condition1'gB} and \eqref{condition2'gB} shows much better their close relation, indicating at the usefulness of treating $g$ and $B$ on more or less the same footing.
This is implemented in \cite{Killing2}.

We summarize our findings in the following two statements:
\begin{prop}
Let $E$ be an anchored bundle over $M$ fitting into the sequence \eqref{FETM}, exact on the level of sections, and the anchor $\rho$ generating the possibly singular foliation ${\cal F}$ on $M$. Then the functional \eqref{S12} provides a gauging of the functional \eqref{S02} along ${\cal F}$ (cf.~Definition \ref{Def1}) if and only if the bundle $E$ can be equipped with a connection $\nabla$ and an  endomorphism-valued 1-form $\phi$ such that the metric $g$ and the 2-form $B$ on $M$ satisfy the following pair of equations: Eqs.~\eqref{condition1'gB} and \eqref{condition2'gB}, which in components take the form of Eqs.~\eqref{condition1gB} and \eqref{condition2gB}, or, equivalently, Eqs.~\eqref{Sym} and \eqref{Alt}, where the upper/lower sign in each of the second equations refers to Euclidean/Lorentzian signature of the world-sheet metric. \label{Prop1}
\end{prop}

\begin{prop} Under the conditions of the previous proposition and equipping $E$ with a compatible almost Lie algebroid structure, the (regular part of the) infinitesimal gauge transformations of \eqref{S12} can be always parametrized as in Eqs.~\eqref{deltaXA} for arbitrary $\epsilon \in \Gamma(X^*E)$ and $\lambda \in \Omega^1(\Sigma,X^*F)$. The resulting space of tangent vectors to the map $a \sim (X,A)$ does not depend on the choice of the almost Lie algebroid structure. \label{Prop2}
\end{prop}
An almost Lie algebroid structure on $E$ compatible with the anchored bundle structure $(E,\rho)$ on $M$ is a choice of a product on its sections such that its structure functions \eqref{structurefunctions} are consistent with the morphism property of $\rho$, i.e.~they also govern the involutivity of the image of $\rho$, Eq.~\eqref{C}. Being given the anchor $\rho$, one thus has an ambiguity in the definition of the product which lies in the kernel of the anchor map. Since we assumed \eqref{FETM} to be exact (on the level of sections, but this is sufficient here), the ambiguity lies in the image of the map $t$. Correspondingly, if one makes another choice for the structure functions, say ${C'}_{ab}^c$, then the difference to the parametrization of $\delta A$ in \eqref{deltaXA} lies in the image of $t$ and can be reabsorbed by an appropriate, field-dependent choice or change of $\lambda$. This proves the last part of Prop.~\ref{Prop2} and shows that the definition of the gauging of \eqref{S02} as given in \eqref{S12} only needs the data mentioned in Proposition \ref{Prop1}, also on the level of the off-shell symmetries. 

To see, if this gauging is strict or some weaker version of it, cf.~Def.~\ref{Def2}, we now need to turn to the analysis of the equations of motion.

\subsection{Euler-Lagrange equations in the case of a smooth quotient}
\label{sec:2.3}

Let us first simplify the notation in the gauged and ungauged theories, Eqs.~\eqref{S12} and \eqref{S02}, respectively. We have found in Prop.~\ref{Prop1} that 
\begin{equation}
S[a] = \frac{1}{2} \int_\Sigma E_{ij}(X)\, \dD X^i \wedge (1+\ast) \dD X^j
\end{equation}
is a gauging of 
\begin{equation}\label{SE}
S_0[X] = \frac{1}{2} \int_\Sigma E_{ij}(X)\, \dd X^i \wedge (1+\ast) \dd X^j
\end{equation}
where $E_{ij} \equiv g_{ij} + B_{ij}$ with $g$ and $B$ satisfying the extended invariance conditions specified in the Proposition \ref{Prop1} and where, as before, $\dD X \equiv \dd X - \rho(A)$. Let us for simplicity of the notation assume from now on that we deal with the ``physical'' Lorentzian signature of $\gamma$ on the world sheet. In this case, we can introduce light-cone coordinates $\sigma^{\pm}$ such that $\gamma = \exp(\nu) \dd \sigma^+ \vee \dd \sigma^-$ for some locally defined function $\nu$ on $\Sigma$. Since then $\ast \dd \sigma^\pm = \mp \dd \sigma^\pm$, the action \eqref{S12} takes the form
\begin{equation}\label{SDE}
S[a]\equiv \int_\Sigma E_{ij}(X) \, D_+ X^i D_- X^j \: \dd \sigma^+ \wedge \dd \sigma^- \, 
\end{equation}
with $D_\pm X^i \equiv \partial_\pm X^i- v_a^i(X) A^a_\pm$. 

In this subsection, we assume that the foliation is regular and has a good quotient. Being regular implies that $\ker\rho$ is a vector subbundle of $E$ and we have the following exact sequence of vector bundles
\begin{equation}\label{ker}
0 \to \ker \rho \to E \to T  {\cal F} \to 0 
\end{equation}
such that we can identify the bundle $F$ in \eqref{FETM} with $\ker \rho$, the map $t$ becoming an embedding. In particular, the quotient of $E$ by $\ker \rho$ is isomorphic to the tangent sub-bundle $T {\cal F} \subset T M$ of the foliation ${\cal F}$. This quotient is taken precisely by the $\lambda$-symmetry of $A$ in Eq.~\eqref{deltaXA}. Let us denote this equivalence class of $A$ with respect to the $\lambda$-symmetry by $[A]$. Then $[A] \cong V \equiv \rho(A)$, i.e.~knowing the $\lambda$-class of $(A^a)_{a=1}^r$ is equivalent to knowing $(V^i)_{i=1}^n \equiv (v_a^i(X) A^a)_{i=1}^n \in \Omega^1(\Sigma,X^*TM)$. In more common physics terms, one might formulate this by choosing a gauge condition fixing the $\lambda$-invariance of the functional $S$. This would correspond to choosing a splitting of the sequence 
\eqref{ker} which then permits to identify $E$ with 
$\ker \rho \oplus T{\cal F}$ and setting the $\ker(\rho)$-part of $A$ to zero---this is a gauge condition which we can implement off-shell and on-shell since this part of the $A$-field does not enter the functional $S$. After such a choice of gauge, one has effectively $A|_{\lambda-gauge-fixed} \cong V$. In any of these cases, making use of the $\lambda$-invariance, we can determine $A$ (i.e.~actually $[A]$ or  $A|_{\lambda-gauge-fixed}$) iff we can determine $V$, i.e.~$V^i_\pm (\sigma) \equiv v_a^i(X(\sigma)) A_\pm^a(\sigma)$.

Let us furthermore choose adapted coordinates $(X^i)=(X^I,X^\alpha)$ on the target, such that setting the first $k$ coordinates $X^I$ to constants precisely determines a leaf of the foliation locally. In this coordinate system, the vector fields $v_a= v_a^\alpha \partial_\alpha$ form an over-complete basis of vector fields tangent to $T{\cal F}$. 

Variation of $S$ with respect to $A_+^a$ yields
$ v_a^\alpha E_{\alpha i} D_- X^i=0$. Due to the over-completeness of the $v_a$s, this is tantamount to $E_{\alpha i} D_- X^i=0$. Likewise the variation of \eqref{SDE} w.r.t.~$A_-^a$ yields $E_{i \alpha}(X) D_+ X^i = 0$. In the adapted coordinate system,  $D_\pm X^I = \partial_\pm X^I$ and $D_\pm X^\alpha = \partial_\pm X^\alpha - V_\pm^\alpha$. Moreover, $E_{\alpha \beta}=g_{\alpha \beta} + B_{\alpha \beta}$ is an invertible matrix due to the non-degeneracy of $g$ restricted to a leaf; we denote the inverse matrix by $(E^{\alpha \beta})$ so that 
$E^{\alpha \beta}E_{\beta\gamma} = \delta^\alpha_\gamma$. In this notation one then easily verifies that the $A_\pm$-variation of \eqref{SDE} found above permit us to express $V_\pm\equiv V_\pm^\alpha \partial_\alpha$ algebraically in terms of the other fields as follows:
\begin{equation}\label{Vpm}
V_\pm^\alpha = \partial_{\pm} X^\alpha + M_{\pm,I}^\alpha \, \partial_\pm X^I
\end{equation}
where we introduced the matrices 
\begin{equation}\label{Mpm}
M_{+,I}^\alpha \equiv E_{I\beta}E^{\beta\alpha} \quad , \qquad M_{-,I}^\alpha \equiv  E^{\alpha\beta} E_{\beta I} \;.
\end{equation}
Up to the $\lambda$-symmetry dealt with already before in one or the other way mentioned, the $A$-variation of $S$ permits us to express $A$, and we may put the Equations \eqref{Vpm} directly into the action without losing any further field equation of the original problem. Thus we may replace $E_{ij} \, D_+X^i D_-X^j$ by $\left(E_{IJ} - E_{I\alpha} M_{- ,J}^\alpha - M_{+,I}^\alpha E_{\alpha J} + M_{+,I}^\alpha E_{\alpha\beta}  M_{- ,J}^\beta\right) \partial_+ X^I \partial_- X^J$. Using the explicit form of the matrices \eqref{Mpm}, the matrix coefficient can be further simplified to
\begin{equation}\label{Eredu}
E^{red}_{IJ} = E_{IJ} - E_{I\alpha}E^{\alpha\beta} E_{\beta J}
\end{equation}
with the simplified or ``reduced'' action taking the form
\begin{equation}\label{Sredu}
S^{red} =  \int_\Sigma E^{red}_{IJ}\: \partial_+ X^I  \partial_- X^J \; \dd \sigma^+ \wedge \dd \sigma^- \, . 
\end{equation}
Since $E_{IJ}$ can be decomposed uniquely into a symmetric part $g_{IJ}$ and an antisymmetric part $B_{IJ}$, strictness of the gauging \eqref{SDE} is established as soon as we have shown that $S^{red}$ depends only on the coordinates $X^I$ parametrizing the leaf-space, but no longer on the coordinates $X^\alpha$. 

This, however, follows from an elegant argument using Noether's second identity for the $\epsilon$-gauge invariance. Denoting $\epsilon^\alpha := v_a^\alpha \epsilon^a$, we see that $\delta_\epsilon X^\alpha = \epsilon^\alpha$, while $\delta_\epsilon X^I = 0$. The second Noether identity now implies that 
\begin{equation}
\int_\Sigma \left( \epsilon^\alpha \frac{\delta S}{\delta X^\alpha} + \delta_\epsilon A^a \wedge \frac{\overrightarrow{\delta} S}{\delta A^a} \right) = 0  
\end{equation}
for arbitrary $\epsilon^\alpha$. On the other hand,
\begin{equation}
S^{red} [X^I,X^\alpha] = S[X^I,X^\alpha,A]|_{\frac{\overrightarrow{\delta} S}{\delta A^a}=0} \quad .
\end{equation}
Using the above two equations together with the chain rule, we indeed obtain $\frac{\delta S^{red}}{\delta X^\alpha} = 0$ or, equivalently, 
\begin{equation}\label{Ered}
E_{IJ}^{red} = E_{IJ}^{red}(X^I) \quad , \qquad S^{red} = S^{red} [X^I] \, ,
\end{equation}
as claimed before. Thus in total we find as a summary of this section
\begin{theorem} \label{Thmred}
The action functional $S[a]$ given by minimal coupling, Eq.~\eqref{S12}, is a strict gauging (outside singularities) of the bosonic string theory $S_{bos}[X]$, Eq.~\eqref{SgB} or Eq.~\eqref{S02}, in the sense of Definition \ref{Def2} (for any foliation and choice of $g$ and $B$ as specified in Proposition \ref{Prop1}).
\end{theorem}
Note that we assumed a regular foliation with a good quotient in our considerations. However, if there are singularities, we can consider $M_{reg}$ and small enough balls $B \subset M_{reg}$ to apply the considerations above (cf.~the text before Definition \ref{Def2}).

A final remark, suppose that $E=g$ and thus $B=0$ in the functionals \eqref{SE} and \eqref{SDE}. Then the formula for the reduced metric is $g^{red}_{IJ} = g_{IJ} - g_{I\alpha} g_{J\beta} g^{\alpha \beta}$ and $M^\a_{+,I} = M^\a_{-,I} =: M^\a_{I}$. On the other hand, one calculates 
\begin{equation} \label{decomp}
g \equiv \sfrac{1}{2} g_{ij} \dd x^i \vee \dd x^j = 
\sfrac{1}{2} g^{red}_{IJ}\dd x^I \vee \dd x^J + \sfrac{1}{2}  g_{\a\beta} (\dd x^\a + M^\a_I \dd X^I) \vee  (\dd x^\beta + M^\beta_J \dd X^J) \, . 
\end{equation}
Since $(\dd x^\a + M^\a_I \dd X^I)$ together with $\dd x^I$ is the orthogonal basis induced by $g$ itself, we see that the decomposition \eqref{decomp} is the one of $g$ into $g_\perp=\sfrac{1}{2} g^{red}_{IJ}\dd x^I \vee \dd x^J$ and $g_\parallel$\footnote{In physics, such a decomposition is familiar also in the context of dimensional reduction of supergravity theories \cite{ScherkSchwarz,MaharanaSchwarz}.}. In \cite{Kotov:2014iha} it was shown that the condition on the existence of $\nabla$ in \eqref{condition1gB}
with $B=0$ is equivalent to ${\cal L}_{v_a} (g_\perp) = 0$ and thus to $g^{red}_{IJ} = g^{red}_{IJ} (X^I)$. While there the proof was a purely geometric one, here we established this identity by gauge invariance of a functional. 
Similarly, the more general first equation in Eq.~\eqref{Ered} was found by purely gauge theoretic considerations. For a purely geometric derivation, using directly the two equations Eq.~\eqref{condition1gB} and Eq.~\eqref{condition2gB}, can be found in \cite{Killing2}. 

For the Euclidean signature of $\gamma$, it is best to introduce complex coordinates $z$ and $\bar{z}$. Then one ends up essentially again with Eq.~\eqref{SDE}, but where $E_{ij} = g_{ij} + i B_{ij}$. While before $(E_{ij})$ was the most general real matrix, now it is the most general hermitian one. Thereafter everything proceeds analogously, arriving at equations of the form \eqref{Eredu}, \eqref{Sredu}, and \eqref{Ered}. It is important to note here that the matrix $E^{red}_{IJ}$ is again hermitian, and thus can be decomposed into $g_{IJ}^{red} + i B_{IJ}^{red}$. 
This establishes Theorem \ref{Thmred} also for Euclidean signature of the world-sheet metric $\gamma$. 

We finally remark here that the exactness of the sequence \eqref{FETM} ensures the $\lambda$-invariance \eqref{deltalambda} of the functional \eqref{S12}, so that Theorem \ref{Thmred} does not come as a big surprise in view of the discussion in Sec.~\ref{Orientation} following the functional \eqref{rot'}.

\section{Beyond Minimal Coupling---\\Universal Form of the Gauged Action with WZ-term}
\label{Sec3}

\subsection{A simple example, first orientation}
\label{Sec:3.1}

As we mentioned already at the end of section 
\ref{Sec:2.1}, there is a simple situation arising already in the context of ordinary gauging of a Lie algebra $\mathfrak{g}$ for the sigma model \eqref{S02} when the minimal coupling \eqref{S12} is not sufficient: Namely when the change of $B$ with respect to a $\mathfrak{g}$-action is not zero but just exact,\footnote{In this section, we will use the notation $\rho_a$ instead of $v_a$; in the general case corresponding to \eqref{FETM}, one then has $\rho(e_a) = \rho_a\equiv v_a$ for any local frame $e_a$ in $E$. In the case of a group or Lie algebra action, one has $E=M\times \mathfrak{g}$ and the $\rho_a$ used below corresponds to a constant frame $e_a$ or, equivalently, to a basis of the Lie algebra $\mathfrak{g}$.} 
\begin{equation} \label{dbeta}
\cL_{\rho_a} B = \dd \beta_a \, . 
\end{equation}
The original functional \eqref{S02} is then invariant with respect to the rigid symmetry with $\epsilon = const$, while \eqref{S12} is not even rigidly invariant, needless to say not gauge invariant. However, in some cases one may still find a gauge invariant extension of the original functional \eqref{S02}. 

Let us look at a simple example when this can be done. Let $X \colon \Sigma \to \R^3$, the flat target being equipped with the standard metric $g = \dd x^2 + \dd y^2 + \dd z^2$ and the following 2-form:
\begin{equation}\label{xdydz}
B = x \dd y \wedge \dd z\: .
\end{equation}
While the metric is strictly invariant with respect to rotations $G=SO(3)$, the above 2-form is so up to an exact form as in Eq.~\eqref{dbeta}. Let us be more explicit here: Denote by $\rho_a = -\varepsilon_{abc} x^b \partial_c$ possible generators of the Lie algebra $\mathfrak{g}=so(3)$, $[\rho_a,\rho_b]=\varepsilon_{abc} \rho_c$. Then a direct calculation yields:
\begin{equation}
\cL_{\rho_x} B = 0 \:, \quad \cL_{\rho_y} B = \dd \left(\sfrac{1}{2}(z^2-x^2) \dd y \right) \:, \quad 
\cL_{\rho_z} B = \dd \left( \sfrac{1}{2}(y^2-x^2) \dd z \right) \: .
\end{equation}
Note that this fixes $\beta_a$ only up to closed  contributions:
\begin{equation}\label{betaxyz}
\beta_x = \dd f_x \: , \quad\beta_y = \sfrac{1}{2}(z^2-x^2)\dd y + \dd f_y\: , \quad\beta_z = \sfrac{1}{2}(y^2-x^2) \dd z + \dd f_z \: ,
\end{equation}
for three arbitrary functions $f_a$. Here we used that in $\R^3$ any closed form is automatically exact. 

Since we assume $\Sigma$ to have no boundary, we can thus add a closed form to $B$, leaving 
\begin{equation}\label{H}
H = \dd B = \dd x \wedge \dd y \wedge \dd z
\end{equation}
unchanged. In the present case, this observation can be used to gauge the model \eqref{S02} for the above $g$ and $B$ in an elegant way. The idea is as follows: Instead of looking at the model defined by $B$ as above, we can gauge another model with an inherently rotation invariant 2-form $B^{inv}$ as long as only the 3-form \eqref{H} is the same. Such an invariant 2-form may not need to exist in general, here however it does:
\begin{equation}\label{Binv}
B^{inv} = \sfrac{1}{3} \left( x \dd y \wedge \dd z + y \dd z \wedge \dd x+z \dd x \wedge \dd y\right)\equiv \sfrac{1}{6} \varepsilon_{ijk} x^i \dd x^j \wedge \dd x^k \: .
\end{equation}
$B^{inv}$ is inherently $so(3)$-invariant, $\cL_{\rho_a} B^{inv} = 0$ for $a=1,2,3$, and it satisfies  $\dd B^{inv} = H$. Thus there is a 1-form $C \in \Omega^1(\R^3)$ such that $B + \dd C = B^{inv}$ and correspondingly, for a boundary-less $\Sigma$, \eqref{S02} is not changed at all when replacing $B$ by $B^{inv}$. For \emph{this} choice of the 2-form, on the other hand, we may apply standard minimal coupling to obtain the gauged action. Thus gauging rotations for the functional 
\begin{equation}\label{S0example}
S_0[X,Y,Z] =  \int_\Sigma \left(\partial_+ X^i \partial_- X^i  + X \partial_+ Y \partial_- Z - X \partial_+ Z \partial_- Y \right) \dd \sigma^+ \dd \sigma^- \, ,
\end{equation}
where $(X^i)_{i=1}^3 \equiv (X,Y,Z)$ and we used 
 light-cone coordinates on the world-sheet for simplicity, is effectuated by
\begin{equation} \label{Binvaction}
S_1[X^i,A^a] =  \int_\Sigma \left(\dD_+ X^i \dD_- X^i  + \sfrac{1}{3} (X \dD_+ Y \dD_- Z - X \dD_+ Z \dD_- Y + cycl{(X,Y,Z)}) \right) \dd \sigma^+ \dd \sigma^- \, ,
\end{equation}
with the usual covariant derivatives as in Eq.~\eqref{DX0}, i.e., e.g., $\dD X = \dd X + Z A^y - Y A^z$. For clarity, using Stokes theorem to get rid of the $C$-contribution to \eqref{Binvaction}, and eliminating contributions that cancel against one another in the cyclic permutations, we may rewrite Eq.~\eqref{Binvaction} also according to
\begin{eqnarray}
S_1[a] &=& \frac{1}{2}\int_\Sigma \, g_{ij} \:\dD X^i \wedge * \dD X^j +  \int_\Sigma X \dd Y \wedge \dd Z \label{S1example}\nonumber  \\
&& +\frac{1}{3}\int_\Sigma A^x \wedge \left( Y^2 \dd X + Z^2 \dd X - XY \dd Y - XZ \dd Z \right)    +  cycl \\ &&+\frac{1}{3}\int_\Sigma X\left(X^2+Y^2+Z^2 \right) A^y \wedge A^z+  cycl\: .\nonumber 
\end{eqnarray}
For later purposes, we note that this gauged action is of the form
\begin{equation}\label{S1form}
S_1[a] = \int_\Sigma \, \sfrac{1}{2}X^*(g_{ij}) \:\dD X^i \wedge * \dD X^j + X^*B + A^a \wedge X^* \alpha_a + \sfrac{1}{2} X^*(\gamma_{ab}) A^a \wedge A^b \, ,
\end{equation}
where $\alpha_a \in \Omega^1(M)$ and $\gamma_{ab} \in C^\infty(M)$, with here $M\equiv \R^3$,  and where for clarity we exceptionally emphasized the pull-back by $X$ explicitly. The action functional \eqref{S1example} corresponds to the following choice of $\alpha$s and $\gamma$s: 
\begin{eqnarray}
\alpha_x &=& \sfrac{1}{3}(y^2 + z^2) \dd x - \sfrac{1}{3}x (y \dd y + z \dd z)\label{alphaxgood}\\
\gamma_{yz}&=&\sfrac{1}{3} x r^2 \, ,
\end{eqnarray}
 where $r = \sqrt{x^2+y^2+z^2}$ is the distance to the origin in the target and the other components of $\alpha_a$ and $\gamma_{ab}$ follow from cyclic permutations and the fact that $\gamma_{ab}=-\gamma_{ba}$. Let us denote these quantities by a superscript ``good''. Their choice, rendering the action \eqref{S1form} an admissible and in some to-be-specified sense also a good gauging of $S_0$, can be written in an elegant form as follows:
\begin{equation}
\alpha_a^{good} = -\iota_{\rho_a} B^{inv} \; , \qquad \gamma_{ab}^{good} = \iota_{\rho_a} \alpha_b^{good} \; . \label{good}
\end{equation}
These expressions permit to immediately deduce the following two important relations:
\begin{eqnarray}
\cL_{\rho_a} \alpha_b &=& C^c_{ab} \alpha_c \label{equiv}\\
\iota_{\rho_a} \alpha_b + \iota_{\rho_b} \alpha_a &=&0 \, , \label{iso}
\end{eqnarray}
where $C^c_{ab}$ denote the structure constants of the Lie algebra $\mathfrak{g}$, which here equal to $\varepsilon_{abc}$. We will come back to these two equations as well as to Eq.~\eqref{talpha} in a more general context repeatedly below, for which reason we omit(ted) the superscripts.

Let us also observe another identity that follows immediately from a choice of the form \eqref{good}: Whenever the generators $\rho_a$ of the Lie algebra $\mathfrak{g}$ are not independent, i.e.~one has $s$ identities of the form $t^a_I \rho_a \equiv 0$, cf.~also our assumption \eqref{FETM},
then automatically
\begin{equation}\label{talpha}
t^a_I \: \alpha_a = 0 \qquad \forall I = 1,\ldots, s.
\end{equation}
In the present context of the SO(3)-invariant theory \eqref{S0example}, the sequence \eqref{FETM} takes the form
\begin{equation}\label{FETMso3}
\R \times M \stackrel{t}{\longrightarrow} so(3) \times M \stackrel{\rho}{\longrightarrow} TM\, ,
\end{equation}
where $M=\R^3$ and also $so(3) \cong \R^3$, $t(1)=(x,y,z)$, and $\rho(e_a) = \rho_a$ as given above. Clearly $t^a \rho_a = -x^a \varepsilon_{abc} x^b \partial_c \equiv 0$ as required. Note also that the sequence \eqref{FETMso3} is pointwise exact except for at the origin in $M=\R^3$, where all the $\rho_a$s vanish identically. However, there is no smooth map $\rho$ supported only at this point in $M$; so the sequence \eqref{FETMso3} is indeed exact at the level of sections. In addition to the equations \eqref{equiv} and \eqref{iso}, the choice \eqref{good} also satisfies
\begin{equation}\label{xalphagood}
x \,\alpha_x^{good} + y\, \alpha_y^{good}+ z\, \alpha_z^{good}\equiv 0 \, .
\end{equation}

Despite the fact that without a boundary, $\partial \Sigma = \emptyset$, the functional with $g$ and $B$ and the one with $g$ and $B^{inv}$ are \emph{equal}, applying minimal coupling to each of them gives inequivalent functionals: In the former case we would obtain
\begin{equation}
\alpha_a^{wrong} = -\iota_{\rho_a} B \; , \qquad \gamma_{ab}^{wrong} = \iota_{\rho_a} \alpha_b^{wrong} \;  \label{wrong}
\end{equation}
in \eqref{S1form}, which then is even not gauge invariant. In fact, given the functional \eqref{S1form}, one may ask under which conditions on $\alpha$s and $\gamma$s this becomes a gauging of \eqref{SgB}---for the standard transformation properties of $\mathfrak{g}$-valued connection 1-forms $A$. It is not difficult to verify that in this case Eqs.~\eqref{equiv} and \eqref{iso} together with 
\begin{equation}\label{dalpha}
\iota_{\rho_a} H = \dd \alpha_a
\end{equation}
is necessary and sufficient for the gauge invariance (under the above conditions, i.e.~in particular that $A$ follows the \emph{usual} gauge transformation rule \eqref{gauge1b}). Here, as before, $H=\dd B$. Rewriting Eq.~\eqref{dalpha} in terms of $B$ and its Lie derivative using $\cL_{\rho_a} = \iota_{\rho_a} \circ \dd + \dd \circ \iota_{\rho_a}$, we find the following possible identification
\begin{equation}\label{alphabeta}
\alpha_a = \beta_a - \iota_{\rho_a}B \, .
\end{equation}
Note however that such as $\beta$s also the $\alpha$s are defined only up to the addition of closed 1-forms, and this choice needs to be taken in an appropriate way \cite{Hull1,Hull2,Figueroa,Plauschinn}:
\begin{lemma}\label{lemmaHull}
Let $\mathfrak{g}$ be part of the isometry Lie algebra of $(M,g)$ and $B$ satisfy Eq.~\eqref{dbeta} with respect to its generators $\rho_a$ for some $\beta_a\in\Omega^1(M)$. Then the functional \eqref{S1form}  provides a gauging of the sigma model \eqref{SgB} with respect to the standard gauge transformations \eqref{gauge1a}, \eqref{gauge1b} if and only if $\beta_a$ can be chosen such that $\alpha_a$, defined in \eqref{alphabeta}, satisfies the two equations  \eqref{equiv} and \eqref{iso} (and $\gamma_{ab} = \iota_{\rho_a}\alpha_b$).
\end{lemma}
That the conditions \eqref{equiv} and \eqref{iso}, together with Eq.~\eqref{dalpha}/\eqref{dbeta}, do not fix the $\alpha$s/$\beta$s uniquely, is nicely illustrated at our example \eqref{S0example}: Evidently $\dd x^i$ is equivariant with respect to the standard so(3)-action on $\R^3$, and it satisfies $\iota_{\rho_a} \dd x^b + \iota_{\rho_b} \dd x^a=0$ for all $a,b \in \{1,2,3\}$. Since it is moreover exact, we may add it to $\alpha_a$ without violating Eq.~\eqref{dalpha}. Thus
\begin{equation}\label{bad}
\alpha_a^{bad} := \alpha_a^{good} + \dd x^a
\end{equation} 
provides another possible gauging according to Lemma \ref{lemmaHull}. 

Actually, instead of the suffixes ``good'' and ``bad'' one could also have used ``strict'' and ``non-strict'': As one may verify by a direct calculation, the field equations of \eqref{S1form} yield the equation $\dd R = 0$ for $R^2 \equiv X^2+Y^2+Z^2$ when using the $\alpha$s from \eqref{bad} (together with $\gamma_{ab}=\iota_{\rho_a}\alpha_b$). Thus the choice \eqref{bad} leads to a freezing of the movement of the string transversal to the orbits of the gauging. 

This is not the case for the choice \eqref{good}: Indeed, one has
\begin{prop} \label{PropHullstrict}
Under the conditions of Lemma \ref{lemmaHull}, the gauging is strict in the sense of Definition \eqref{Def2} iff the choice of $\alpha$s satisfies Eq.~\eqref{talpha}. 
\end{prop} 
Needless to say, that $\alpha^{bad}_a$ does not satisfy  Eq.~\eqref{talpha}, which here takes the form of Eq.~\eqref{xalphagood}, due to $x^i\dd x^i \neq 0$ for $x\neq0$.
We will prove the statement of Proposition \ref{PropHullstrict} in a more general context in the subsequent subsection, cf.~Proposition \ref{Prop4} below. As a simple consequence of it we now have 
\begin{lemma}\label{lemmaBinv}
Whenever one finds a strictly $\mathfrak{g}$-invariant 2-form $B^{inv}$ such that 
$\dd B^{inv} = \dd B \equiv H$, then the functional \eqref{S1form} with \eqref{good} provides a strict gauging of \eqref{SgB}.
\end{lemma}
This statement follows also without using Proposition \ref{PropHullstrict} from Lemma \ref{lemmaHull} together with Theorem \ref{Thmred} by noting that the field equations of the functionals \eqref{SgB} and \eqref{S1form} depend on $B$ only through $H=\dd B$ and that for $B^{inv}$ the conditions of Theorem \ref{Thmred} are satisfied. 

It is illustrative at this point to return to the initial problem of this subsection, posed by the 2-form \eqref{xdydz} and the general solution \eqref{betaxyz} for the 1-forms $\beta_a$. First, since minimal coupling corresponds precisely to \eqref{wrong}, we note a contradiction of this equation with Eq.~\eqref{alphabeta} for non-closed 
$\beta$s---which is evidently not the case for \eqref{xdydz}, \eqref{betaxyz}. The ``good solution'' \eqref{alphaxgood} or \eqref{good}, on the other hand, corresponds to the following choice of the three functions $f_a$ entering the general solution \eqref{betaxyz}:
\begin{eqnarray}
f_x &=&  \sfrac{1}{3} x(y^2+z^2)\: , \nonumber \\
f_y &=& -\sfrac{1}{6} y(x^2+z^2)\: , \label{fxyz}\\
f_z &=&  -\sfrac{1}{6} z(x^2+y^2)\: . \nonumber 
\end{eqnarray}
The choice \eqref{bad}, which also provides a gauging according to Lemma \ref{lemmaHull}, but not a strict one, results from Eqs.~\eqref{fxyz} by the replacement $f_x \to f_x + x$, $f_y \to f_y + y$, $f_z \to f_z + z$.

Starting from the general ansatz/solution \eqref{betaxyz} for $\beta$s or $\alpha$s, where their relation is determined by Eq.~\eqref{alphabeta}, the conditions \eqref{equiv} and \eqref{iso} for gauge invariance reduce to coupled second order differential equations on the three functions $f_a$; the additional condition \eqref{xalphagood} for a strict gauging adds a further (first order) differential equation to this. 
In view of this coupled system of differential equations, the  passage by Lemma 
\ref{lemmaBinv}---together with the rather obvious choice for $B^{inv}$ given in Eq.~\eqref{Binv}--- strikingly simplified the problem.

\subsection{General ansatz for the gauging of a foliation \\ for the 2d standard sigma model with WZ-term}
\label{Sec:3.2}

The main purpose of this section is to generalize the discussion of the previous subsection to  gauging of foliations. But note that even if one restricts to a foliation that results from a group action, the discussion above can be generalized since we restricted the $A$-gauge transformations unnecessarily to be of the form \eqref{gauge1b}: The strict invariance of $g$ and $B$ can be generalized to \eqref{condition1gB} and \eqref{condition2gB} even in the case of a $\mathfrak{g}$-action if one just relaxes the transformation properties of the $A$-field. Similarly, we now want to study to what strict invariance of $g$ and relative invariance \eqref{dbeta} of $B$  with respect to a $\mathfrak{g}$-action generalizes to if we consider arbitrary foliations on $M$ (fitting into the sequence \eqref{FETM}, as always in this paper) and do not further restrict the transformation behavior of the gauge field $A$. 

In the discussion above we also noticed that in the present context it is mainly the ``curvature'' $H=\dd B$ of the ``gerbe connection'' 2-form $B$ that enters the considerations for gauging the sigma model in the most general setting.  We  essentially have many \emph{different} functionals of the form \eqref{SgB} parametrized by different choices of $B$ that we will be able to gauge by the \emph{same} choice of $\alpha$s and $\gamma$s by an action of the form \eqref{S1form} provided only that $H$ is the same for them. Moreover, also the field equations resulting from \eqref{SgB} depend on $B$ only through its exterior derivative $H$. We thus may rewrite the functional \eqref{SgB} to be gauged \emph{symbolically} in the form 
\begin{equation} \label{S0WZ}
S_{0WZ}[X] = \int_{\S}\sfrac 12 g_{ij}(X)\dd X^i\w\ast \dd X^j + \int H \, .
\end{equation}
In fact, we will consider the problem even for the case that $H$ is only closed and not necessarily exact. For closed forms $H$ with integer cohomology, one may make sense of \eqref{S0WZ} as a multi-valued functional or, better, 
as a single-valued contribution $\exp(\frac{i}{\hbar} \lambda \int H)$ to the path integral for appropriate choices of 
the coupling constant $\lambda$ (cf., e.g., \cite{WZW}); but this is not the main focus of the present paper and we will content 
ourselves here with the Euler-Lagrange equations mainly, which are perfectly well-defined for closed $H$. Likewise, also gauging makes sense at the level of differential equations already. 

Guided by the previous subsection and Eq.~\eqref{S1form}, we will content ourselves also here with minimal coupling for the $g$-part of the action and consider an ansatz for the gauging of \eqref{S0WZ} in the form of
\begin{equation} \label{SWZ}
S_{1WZ}[X,A]=\frac{1}{2}\int_\Sigma g_{ij} \dD X^i \w \ast \dD X^j + \int H + \int_\Sigma A^a \wedge \alpha_a + \frac{1}{2} \gamma_{ab} A^a \wedge A^b \: , 
\end{equation}
where, as before, $\alpha_a$ are 1-forms and $\gamma_{ab}$ functions on $M$, both pulled back by $X \colon \Sigma \to M$.   
Likewise, we will consider gauge transformations of the form 
\bea 
\label{depXA} \d_{\epsilon}X^i&=&\rho^i_a(X)\epsilon^a~,\\
\nonumber
\d_{\epsilon} A^a&=&\dd \epsilon^a+C^a_{bc}(X)A^b\epsilon^c+ \omega^a_{bi}(X)\epsilon^b \, \dD X^i + \phi^a_{bi}(X) \epsilon^b \ast \!\dD X^i~.
\eea
for yet undetermined coefficients $\omega^a_{bi}$ and $\phi^a_{bi}$. This is a fairly general ansatz and also covers all that we found in the previous section. 

Certainly, one may consider a more general ansatz for the functional \eqref{SWZ} and the symmetries \eqref{depXA}. For example,
one could leave the coefficients of terms of the form $A \w \ast \dD X$ and $A \wedge \ast A$ in \eqref{SWZ} open; in the above their coefficients are fixed by minimal coupling to $g$. Or we could add further terms to \eqref{depXA}, such as one proportional to $\ast A$ or $\ast \dd \epsilon$, etc. The calculation then becomes much more technical and the result less transparent. For this reason such a generalization is deferred   to Appendix \ref{AppendixA}.

\begin{prop}\label{Prop3}
The functional \eqref{SWZ} is invariant with respect to gauge transformations of the form \eqref{depXA} if and only if the following equations hold true: The metric $g$ and the closed 3-form $H$  satisfy
\bea 
\label{condition1gH} {\cal L}_{{\rho}_a}g&=&\o^b_a\vee\iota_{{\rho}_b}g+\phi^b_a\vee \alpha_b~,
\\
\label{condition2gH} \iota_{\rho_a} H &=&\dd \alpha_a - \o^b_a\w\alpha_b \pm \phi^b_a\w \iota_{{\rho}_b}g~.
\eea 
In addition,  $\gamma_{ab} = \iota_{\rho_a} \alpha_b$, the antisymmetry of which yields the consistency condition $\iota_{\rho_a} \alpha_b +  
\iota_{\rho_b} \alpha_a = 0$, and finally
\begin{equation}\label{Calpha}
{\cal L}_{\rho_a} \alpha_b = C^c_{ab} \alpha_c + \iota_{\rho_b} \left( \dd \alpha_a - \iota_{\rho_a}H\right) \, .
\end{equation} 
\end{prop}
Before turning to its proof, we remark that the above statement reproduces the result of Proposition \ref{Prop1} in the special case where $H=\dd B$ and $\alpha_a = -\iota_{\rho_a} B$: Indeed, then the functional \eqref{SWZ} reduces simply to \eqref{S12}, Equations \eqref{condition1gH} and \eqref{condition2gH}
turn into \eqref{condition1gB} and  \eqref{condition2gB}, respectively, and the final  Equation \eqref{Calpha} becomes tautologically satisfied as one may verify by an explicit calculation.

\textbf{Proof:} The proof follows from direct calculations. We vary the action functional  \eqref{SWZ}  with respect to gauge transformations  \eqref{depXA}. First notice that
\bea
\delta_\epsilon\dD X^i&=&\epsilon^a\left[\left(\rho^i_{a,j}-\rho^i_b\omega^b_{aj}\right)\dD X^j-\rho^i_b\phi^b_{aj}\ast \!\dD X^j\right]~,
\\
\delta_\epsilon\alpha_b&=&\epsilon^a{\cal  L}_{\rho_a}\alpha_b+(\iota_{\rho_a}\alpha_b)\dd\epsilon^a~,
\eea
and for $H$ closed and $\Sigma=\partial\Sigma_3$
\beq
\delta_\epsilon\int_{\Sigma_3} H=\int_\Sigma\epsilon^a\iota_{\rho_a} H. 
\eeq
Collecting terms appropriately, we then find 
\bea\label{var S}
\delta_\epsilon S_{\text{WZ}} = \int_{\S} \epsilon^a \hspace{-0.5cm}&& \left\{
\sfrac 12\left({\cal L}_{{\rho}_{a}}g-\omega^b_a\vee\iota_{\rho_b}g-\phi^b_a\vee\theta_b\right)_{ij}\dD X^i\w\ast\dD X^j\right.+\nn\\
 &&+\sfrac 12
\left( \iota_{\rho_a}H-\dd\theta_a+\omega^b_a\w\theta_b\mp\phi^b_a\w\iota_{\rho_b}g\right)_{ij}\dD X^i\w \dD X^j+\nn\\
&&- \left( {\cal L}_{\rho_a} \theta_b - C^c_{ab} \theta_c +\iota_{\rho_b} (\iota_{\rho_a}H-\dd\theta_a )+(\gamma_{bd}-\iota_{\rho_b}\theta_d 
)\omega^d_a\right)_i \dd X^i\w A^b+\nn\\
&&\left.+\left(\sfrac 12{\cal L}_{\rho_a}\gamma_{bc}+C^d_{ab}\gamma_{cd}-\sfrac 12\iota_{\rho_c}\iota_{\rho_b}(\iota_{\rho_a} H
-\dd\theta_a)+(\gamma_{bd}-\iota_{\rho_b}\theta_d)\iota_{\rho_c}\o^d_a\right)A^b\w A^c\right\}+\nn\\
+\int_{\S} \hspace{-0.5cm}&&(\gamma_{ab}-\iota_{\rho_a}\theta_b)(\dd\epsilon^a\w A^b+\epsilon^c\phi^b_{ci}\dD X^i\w\ast A^a)~.
\eea
Evidently the vanishing of each of the lines of the right hand side is necessary and sufficient for $\delta_\epsilon S =0$. Noting that the third and fifth line together imply the fourth one, completes the proof. $\square$

The geometric, frame-independent significance of the two equations \eqref{condition1gH} and \eqref{condition2gH} can be 
found as in Section \ref{Sec:2.2}, cf.~Eqs.~\eqref{Sym} and \eqref{Alt}. 
This is due to the fact that, as before, the geometry is already largely determined by the gauge transformations \eqref{depXA}. Here one obtains
\begin{eqnarray}
\mathrm{Sym} \nabla \bar{\rho} &=& \mathrm{Sym} \langle \phi \stackrel{\otimes}{,} \alpha \rangle  \, , \label{barrholambda}\\
\iota_\rho H &=& \dD \alpha \pm \langle \phi \stackrel{\wedge}{,} \bar{\rho} \rangle  \, , \label{iotaHlambda}
\end{eqnarray}
where $\bar{\rho} \equiv \iota_{\rho_a} g \otimes e^a$ and  $\alpha \equiv \alpha_a \otimes e^a$ both take values in  $\Gamma(T^*M \otimes E^*)$, $\phi \in \Gamma(T^*M \otimes E^* \otimes E)$. 
 
Proposition \ref{Prop3} provides the conditions for gauging \eqref{S0WZ} along a foliation, but gives no information on the
fact if the gauging is strict or not. This is addressed in the following
\begin{prop}\label{Prop4}
Under the conditions mentioned in Prop.~\ref{Prop3}, the gauging is strict, if and only if also Equation \eqref{talpha} 
holds true or, equivalently, if and only if the functional \eqref{SWZ} is also invariant with respect to the
$\lambda$-translations \eqref{deltalambda}.
\end{prop}
\textbf{Proof:} The equivalence of the statement follows directly from the subsequent equation, which is obtained by gauge transforming $X^i$ and $A^a$ with respect to $\epsilon$ and $\lambda$:
\be \label{elinv}
\d_{(\epsilon,\lambda)}S_{1WZ}=\int_{\S}\d_{\l}A^a\w\a_a=\int_{\S}\lambda^I\w t^a_I\a_a~.
\ee
This holds true due to the equations ensuring $\epsilon$-invariance as well as that all the rest of the terms resulting from the $\lambda$-transformations vanish identically due to the fundamental assumption that $t(\rho)=0$; in particular, here we used that $\gamma_{ab} = \iota_{\rho_a} \alpha_b = - \iota_{\rho_b} \alpha_a$. 

Next we show that necessity for strictness: Suppose  that \eqref{elinv} does not vanish identically for all choices of $\lambda$, which implies that $X^*(t^a_I \alpha_a)$ does not vanish identically. Thus the field equations of $S_{1WZ}$ contain in particular the following equations
\begin{equation}\label{necessary}
X^*\left(t^a_I \alpha_a\right) = 0 \, ,
\end{equation}
where we once wrote the pullback by $X$ explicitly for clarity. These equations are non-empty for at least one value of $I$. On the other hand, this enforces transversal freezing, if we can show that the contraction of $t^b_I \alpha_b$ with all $\rho_a$ vanishes (here no pullback). This is indeed the case, since $t^b_I \iota_{\rho_a} \alpha_b = -t^b_I \iota_{\rho_b} \alpha_a \equiv 0$ due to $\rho \circ t \equiv 0$.  

To prove sufficiency of the condition, we first choose an adapted frame and adapted coordinates. Let $b_M$ be a frame in $F$ and thus $e_M := t(b_M)$ spans $\ker \rho \subset E$. We complement this by the choice of some $e_\mu$ to a basis $(e_a) = (e_M,e_\mu)$ in $E$. As before we choose coordinates $(X^i) = (X^I,X^\alpha)$ adapted to the foliation, $X^I=const$ labelling the leaves locally. We may also assume that we restrict to a sufficiently small region in $\Sigma$ such that its image with respect to a fixed $X$ lies inside a region $R \subset M$ such that $H=\dd B$. 

Under these assumptions, $t^a_M \alpha_a = 0$ turns into simply $\alpha_M=0$, which in turn implies that the only non-zero components of $\gamma_{ab}$ are $\gamma_{\mu \nu}$. Also we then may deduce that $\rho^I_\mu = 0 = \rho_M^i$ and that 
$(\rho^\alpha_\mu)$ is an invertible matrix. Then \eqref{SWZ} reduces to
 \begin{equation} \label{SWZ'}
S_{1WZ}[X,A]=\frac{1}{2}\int_\Sigma g_{ij} \dD X^i \w \ast \dD X^j + B_{ij} \dd X^i \wedge \dd X^j +  A^\mu \wedge \alpha_\mu + \frac{1}{2} \gamma_{\mu \nu} A^\mu \wedge A^\nu \: . 
\end{equation}
Since $\dD X^\alpha = \dd X^\alpha - \rho_\mu^\alpha A^\mu$ and $\dD X^I = \dd X^I$, this functional is now explicitly independent of $A^M$. 

We are now going to only show that the $A^\mu$ variation permits to eliminate these fields algebraically, determining it uniquely in terms of the other fields. We will not perform this proof as explicitly as the previous special case analyzed in Sec.~\ref{sec:2.3}. In particular, we will not determine the effective geometry, responsible for the reduced theory here, which we expect to be related to the reduction of Courant algebroids and which we leave as an interesting open problem for other work. Since the fields $X^\alpha$ are completely gauge, and thus can be put to zero, for example, and the fields $X^I$ are gauge invariant, this will complete the proof. All the more as that we already proved that the $X^I$ coordinates are not frozen due to some $A$-component equation that would serve effectively as a Lagrange multiplier for $\dd X^I$. 

To complete the proof, we observe that the variation with respect to $A^\mu$ yields to the following equation:
\begin{equation}
O_{\mu \nu} A^\nu = (\rho^\alpha_\mu g_{\alpha i} * - \alpha_{\mu i}) \dd X^i \, , 
\end{equation}
where the matrix-operator $O_{\mu \nu}$ has the form
\begin{equation}
O_{\mu \nu} \equiv \rho^\alpha_\mu g_{\alpha \beta} \rho^\beta_\nu * + \gamma_{\mu \nu} \, .
\end{equation}
The first, symmetric part of $O_{\mu \nu}$ is evidently non-degenerate and since $\gamma_{\mu \nu}$ is anti-symmetric, so is all of $(O_{\mu\nu})$, which completes the proof. 
$\square$

We summarize our findings up to here in the following  
\begin{theorem}\label{Theorem1}
Let $(E,\rho \colon E \to TM)$ be an involutive anchored vector bundle over $M$. Then  the image of the map $\rho$ integrates to a (possibly singular) foliation ${\cal F}$ of $M$ and $E$ can be equipped with a compatible almost Lie algebroid structure. Consider a variational problem with symbolic action functional \eqref{S0WZ} over an orientable 2-manifold $\Sigma$ without boundary and with target manifold $M$. There exists a local lift of the variational problem $S_{0WZ}[X]$ to a variational problem $S_{1WZ}[a]$, where $a \colon T\Sigma \to E$ and $S_{1WZ}[X,0] = S_{0WZ}[X]$, gauging ${\cal F}$, 
\emph{if} there exists a connection $\nabla$ in $E$ and 1-forms  $\phi \in \Omega^1(M, \mathrm{End}E)$, $\alpha \in \Omega^1(M,E)$ with values in $\mathrm{End}E$ and $E$, respectively, such that $g$ and $H$ satisfy the Equations \eqref{barrholambda} and \eqref{iotaHlambda}---or, equivalently, the component Eqs.~\eqref{condition1gH}, \eqref{condition2gH}---$\iota_{v_a} \alpha_b$ is antisymmetric in $a$ and $b$, and Eq.~\eqref{Calpha} holds true.   
The gauging is strict, if in addition Eq.~\eqref{talpha} holds. In the latter case, the complete generating set of non-trivial and non-singular gauge transformations of $S_{1WZ}$, Eq.~\eqref{SWZ}, is given by the Equations \eqref{deltaXA}. 
\end{theorem}
The statement becomes ``if and only if'' provided one stays within the class of functionals and gauge transformations of 
the form \eqref{SWZ} and \eqref{depXA}, respectively. For a more general ansatz cf.~Appendix \ref{AppendixA}.




\subsection{Universal bosonic sigma model for the gauging in $d=2$}
\label{Sec:3.4}

In \cite{KSS} it was observed that for any ordinary gauging there is a universal gauge theory through which the gauged action factors.\footnote{The observation that gauging of Lie algebras in the presence of a Wess-Zumino term $H$ in two dimensions  is related to Dirac geometry  goes already back to \cite{Anton1}. Further developments and generalizations of this idea were performed e.g.~in \cite{Alekseev-Strobl, Alekseev-Severa}; \cite{KSS} provides the Lagrangian counterpart.} It turns out that this fact goes much further and also covers the gauging of singular foliations. We now recall the essential ingredients for the construction of this universal functional, referring to the literature for further details on Dirac geometry (cf., e.g., \cite{DSM} and \cite{KSS} for some of it).

The universal functional is given for any choice of $(\Sigma,\gamma)$ and $(M,g,H)$, where, as before, 
$\Sigma$ is an orientable surface, preferably without boundary, and $\gamma$ a Riemannian or Lorentzian metric on it, $M$ is an $n$-dimensional manifold equipped with a Riemannian metric $g$ and a closed 3-form $H$. The functional lives on the space of vector bundle morphisms $u$
\begin{equation}\label{u}
u \in \mathrm{Mor}(T\Sigma,TM \oplus T^*M) \quad \leftrightarrow \quad X \in C^\infty(\Sigma,M), \quad V \oplus W \in \Omega^1(\Sigma, X^*TM \oplus X^*T^*M) \: 
\end{equation}
and has the following form
\begin{equation}
S_{univ}[u] = \label{Suniv}
\frac{1}{2}\int_\Sigma g_{ij} {\cal D} X^i \w \ast {\cal D} X^j + \int H + \int_\Sigma W_i \wedge (\dd X^i - \sfrac{1}{2} V^i ) \: , 
\end{equation}
where
\begin{equation}\label{cDX}
{\cal D} X^i \equiv \dd X^i - V^i \: .
\end{equation}
For what follows it will be important to recall two canonical structures that the "generalized tangent bundle" is equipped with \cite{Severa-Weinstein}:
First, there is a bracket on its sections generalizing the Lie bracket of vector fields,
\begin{equation} \label{Courant}
[v\oplus \omega , v' \oplus \omega'] = [v,v'] \oplus \cL_v \omega' - \iota_{v'} \dd \omega - \iota_v \iota_{v'} H   \: .
\end{equation}
Here $v,v' \in \Gamma(TM)$ are vector fields and $\omega,\omega' \in \Gamma(T^*M)$ 1-forms on $M$.
Second, there is a canonical non-degenerate inner product 
\begin{equation} \label{inner}
(v\oplus \omega , v' \oplus \omega') = \iota_v \omega' + \iota_{v'} \omega
 \end{equation}
of signature $(n,n)$. The bracket \eqref{Courant} is called the Courant-Dorfmann bracket \cite{Dorfmann,Courant} and the bundle $TM \oplus T^*M$ the $H$-twisted standard Courant algebroid \cite{lwx,Severa-Letters}. Now we are in the position to formulate the main two theorems of the present paper:
\begin{theorem}\label{Thmmain1}
A gauging of the sigma model \eqref{S0WZ} along a singular foliation $\cF$ on the target manifold $M$ generated by \eqref{FETM} exists, at least if restricted to the form \eqref{S1form} with \eqref{depXA}, \emph{if and only if} there exists a lift $\sigma\equiv (\rho,\alpha)$ of \eqref{FETM},
\begin{equation}\label{FET*M}
 \xymatrix{& & \:\,TM\oplus T^{\ast}M \ar[d]
  \\
    F \ar[r]^{\mathlarger{t}} & E \ar[r]^{\mathlarger{\;\,\rho}} \ar@{-->}[ru]^{\mathlarger{\sigma}}  & TM}
\end{equation} 
with isotropic and involutive image, and the anchored bundle $E$ can be equipped with a connection $\nabla$ and a 1-form-valued endomorphism $\phi$ such that $\sigma$, viewed as the pair of sections $\rho \in \Gamma(TM \otimes E)$ and $\alpha \in \Gamma(T^*M \otimes E)$, satisfies the compatibility conditions Eqs.~\eqref{barrholambda} and \eqref{iotaHlambda} with $g$ and $H$, where $\bar{\rho}\equiv \langle \rho , g \rangle \in \Gamma(T^*M \otimes E)$ and the upper sign in the second equation refers to   Euclidean  signature of the metric $\gamma$ on $\Sigma$, the  lower sign to Lorentzian signature.
The gauging is strict in the sense of Definition \ref{Def2} if and only if also the lifted sequence \eqref{FET*M} is exact (on the level of sections).
\end{theorem}
A maximally isotropic, involutive subbundle $D$ of $TM \oplus T^*M$  is called a Dirac structure. Here isotropy refers to the inner product Eq.~\eqref{inner} and involutivity means that $[\Gamma(D),\Gamma(D)]\subset \Gamma(D)$. If $D$ is just an arbitrary isotropic, involutive subbundle, we call it a small Dirac structure. One may want to extend this latter notion to the case of an involutive and isotropic sub-sheaf of the sections of $\Gamma(TM\oplus T^*M)$. In this sense one could reformulate ``with isotropic and involutive image'' by ``with a small Dirac structure as image''. It is important to note, however, that the image of $\sigma$ inside $T_xM \oplus T_x^*M$ can have a varying dimension with $x \in M$.

{\bf Proof:} Much of the above theorem consists in a reinterpretation of results from before. We still have to check the equivalence of some of the formulas: First, by $C^\infty(M)$-linearity of $\sigma$ and the inner product $(\cdot,\cdot)$, we observe that isotropy of the image holds true if and only if for a basis of sections $e_a$ in $E$ one has $(\sigma(e_a),\sigma(e_a))=0$; with $\sigma(e_a)=\rho_a \oplus \alpha_a$ and Eq.~\eqref{inner} this turns into $\iota_{\rho_a} \alpha_a=0$ or, equivalently, $\iota_{\rho_a} \alpha_b$ anti-symmetric in $a$ and $b$. Second, the image of $\sigma$ is involutive with respect to the Courant-Dorfmann bracket $[ \cdot, \cdot ]$ if and only if it is so for the image of local bases of $\Gamma(E)$; this follows among others from the Leibniz property $[fs,f's'] = ff'[s,s']+f \left(\rho(s)f'\right) s' - f' \left(\rho(s')f\right) s$ valid for arbitrary functions $f, f'$ and sections $s,s' \in \Gamma(TM\oplus T^*M)$ satisfying $(s,s')=0$---this last property is essential here, but may be assumed to hold true due to the isotropy we established already before. By Eq.~\eqref{Courant} we then have $[\sigma(e_a),\sigma(e_b)] = C^c_{ab} \rho_c \oplus \left(\cL_{\rho_a}\alpha_b - \iota_{\rho_b}(\dd \alpha_a - \iota_{\rho_b} H)\right)$. This lies again in the image of $\sigma$ iff the $T^*M$-part of the right-hand side can be written as $C^c_{ab} \alpha_c$. This, however, is precisely the content of Eq.~\eqref{Calpha}. It remains to show that Eq.~\eqref{talpha} is equivalent to the 
exactness of the sequence \begin{equation} \label{sequence}
\Gamma(F)\stackrel{t}{\longrightarrow} \Gamma(E) \stackrel{\rho\oplus \alpha}{\longrightarrow} \Gamma(TM\oplus T^*M).
\end{equation}
If Eq.~\eqref{talpha} is violated, then the image of $t$ does no longer lie in the kernel of $\alpha \colon E\to T^*M$ and thus also not in the kernel of $\rho \oplus \alpha$ in Eq.~\eqref{sequence}. Now the reverse direction: We first observe that Eq.~\eqref{talpha} ensures that the image of $t$ is in the kernel of $\alpha$,  $\mathrm{im}(t)\subset \ker \alpha$. Recall now that $\Gamma(F) 
\stackrel{t}{\longrightarrow} \Gamma(E) \stackrel{\rho}{\longrightarrow} \Gamma(TM)$ is exact by assumption (in any case). Thus $\ker \rho = \mathrm{im}(t)$. Together this implies, $\ker (\rho \oplus \alpha) = \ker \rho \cap \ker \alpha =  \mathrm{im}(t)$.
$\square$

It is worth mentioning that in Theorem \ref{Thmmain1} the choice of the almost Lie algebroid bracket on $\Gamma(E)$, entering for example Eq.~\eqref{deltaXA} through the structure functions, $C^c_{ab} e_c=[e_a,e_b]$, dropped out  
completely. It will be interesting to reinterpret also the conditions Eq.~\eqref{barrholambda} and Eq.~\eqref{iotaHlambda} in terms of generalized geometry, such as it is done with the more special conditions Eq.~\eqref{Sym} and Eq.~\eqref{Alt} in \cite{Killing2}. This, however, is a purely geometrical question surpassing the analysis of the construction of the gauge theories describing strings propagating in quotient spaces and shall be pursued elsewhere.

\begin{theorem}\label{Thmmain2}
Under the conditions of Theorem \ref{Thmmain1}, the action functional $S_{1WZ}$ of the gauged theory is the pullback by $\widehat{\sigma} \colon \mathrm{Mor}(T\Sigma,E) \to \mathrm{Mor}(T\Sigma,TM\oplus T^*M), a \mapsto \sigma \circ a$ of the universal functional \eqref{Suniv}, 
\begin{equation}\label{pullback}
S_{1WZ} = \widehat{\sigma}^*  S_{univ} \: .
\end{equation}
For strict gauging, a complete set of generators of infinitesimal gauge transformations of $S_{1WZ}$ is given by Eq.~\eqref{deltaXA}, with $\epsilon \in \Gamma(X^*E)$, $\lambda \in \Gamma(T^*\Sigma \otimes X^*F)$ arbitrary, for non-strict gauging this holds true with all those $\lambda$ which satisfy $\lambda^I \wedge t^a_I\alpha_a=0$. In the former case, the set of generators does not depend on the almost Lie bracket on $\Gamma(E)$. 
\end{theorem}
{\bf Proof:} We first observe that Eq.~\eqref{pullback} is equivalent to 
\begin{equation}
S_{1WZ}[a]=S_{univ}[\sigma \circ a]\,.
\end{equation} 
Clearly, the base map $X \colon \Sigma \to M$ of $u:=\sigma \circ a$ and $a$ is the same. Thus it remains to relate the $X^* E$-valued 1-forms $A$ with the fields $V$, $W$ in Eq.~\eqref{u} for  the above composed map $u$. Evidently this yields
\begin{equation}\label{VW}
V^i = X^*(\rho_a^i) \, A^a \quad , \qquad  W_i = X^*(\alpha_{ai}) \, A^a \: .
\end{equation} 
Using these expressions to replace $V$ and $W$ in the universal action \eqref{Suniv} together with  Eq.~\eqref{cDX}, the action takes indeed the form of Eq.~\eqref{S1form} if one observes that there, for gauging, it was necessary that $\gamma_{ab}=\iota_{\rho_a} \alpha_b \equiv \rho_a^i \alpha_{bi}$. The remaining part follows from Theorem \ref{Theorem1}, Proposition \ref{Prop2}, and Eq.~\eqref{elinv}. $\square$

The action functional \eqref{Suniv} has been introduced already in \cite{DSM}, with the additional constraint that $u$ has to take values in a (full) Dirac structure. It then becomes what is called the Dirac sigma model and has been constructed so as  to generalize simultaneously the Poisson sigma model \cite{Schaller-Strobl,Ikeda} and the G/G WZW model, Eq.~\eqref{G/H} below with $K=G$. It was shown in \cite{DSM} that the Dirac sigma model is topological, thus not containing any propagating modes. This implies in particular that whenever the anchor map $\rho \colon E \to TM$ is not almost everywhere surjective, but $\sigma  \colon E \to TM \oplus T^*M$ is maximally isotropic, i.e.~spanning a full Dirac structure, the gauging cannot be strict. 
Moreover, there were \emph{no} restricting conditions on the metric $g$ for this to stay true. 
This has an interesting and somewhat surprising geometrical consequence in the case of Dirac structures:
\begin{prop}\label{D}
For $E := D \subset TM \oplus T^*M$ a Dirac structure w.r.t.~the $H$-twisted Courant bracket, there always exist connections $\nabla^\pm = \nabla \pm \phi$ such that for any choice of $g$ and $H$ the gauging exists, i.e.~a lift $\sigma\equiv \rho \oplus \alpha$ in \eqref{FET*M} exists such that the Equations \eqref{barrholambda} and \eqref{iotaHlambda} hold true. The resulting gauged sigma model is topological then and, except for surjective $\rho$, this gauging thus maximally non-strict.
\end{prop}
This Proposition is proven directly in another paper \cite{DSMfourofus}, simultanesouly filling the gap of providing a proof of the 
form of the gauge symmetries of the Dirac sigma model, spelled out in \cite{DSM} without proof or explanation.

If one calls involutive isotropic subsheafs of $\Gamma(TM\oplus T^*M)$ small Dirac structures, it makes sense to consider \eqref{Suniv} with the constraint that $u$ takes values in such a small Dirac structure as a non-topological Dirac sigma model \cite{KSS}---at least if this sheaf has constant rank and can be expressed as sections of a sub-bundle. On the other hand, a priori its fields take values in the generalized tangent bundle $TM \oplus T^*M$ and we content us here to note that it is universal in a certain sense.

We conclude this section illustrating this property of universality. Fix an original functional \eqref{S0WZ}, i.e.~choose $(\Sigma,\gamma)$ as well as $(M,g,H)$. Then for any singular foliation $\cal F$ on $M$ such that the conditions of Eq.~\eqref{FET*M} can be met and any choice of a possible gauged action $S_{gauged}$ for it, strictly or non-strictly gauged, there is a unique map $\sigma$ in Eq.~\eqref{FET*M} giving rise to a (unique) map $\widehat{\sigma} \colon \cA \to \cU, a \mapsto \sigma \circ a$ such that the diagram \eqref{diagrammain} of the Introduction is commutative.
There $\cA \equiv \mathrm{Mor}(T\Sigma,E)$ is the space of fields $(X,A)$ corresponding to the morphism $a\in \cA$ of $S_{gauged}$, $\cU \equiv \mathrm{Mor}(T\Sigma,TM\oplus T^*M)$ the space of fields $(X,V\oplus W)$ of the universal functional $S_{univ}$ and, for simplicity, we considered them as true functionals, i.e.~maps from the space of fields to real numbers. Strictly speaking this is true only if $H=\dd B$ and $\int H$ is understood as the ordinary integral of $X^*B$ over $\Sigma$. But, for example, if $H$ has integer cohomology (appropriately rescaled like, for example, in Eq.~\eqref{WZW}), one can replace the maps $S_{\cdot}$ by $\exp\left(\frac{i}{\hbar} S_{\cdot}\right)$ and $\R$ by $\C$. 

\section{Examples}
\label{Examples}
In this final section, we first want to study some simple examples of the use of the Theorems \ref{Thmmain1} and \ref{Thmmain2}. The emphasis is not on sophisticated examples, but just the simplest ones for an orientation of different types of gaugings and their relation to Dirac structures. It is just to save the reader some time when trying to understand the general, somewhat abstract formalism in the body of this article. The more intricate question of the existence or non-existence of a strict gauging for the SU(2) WZW model, on the other hand, will be addressed in the second part of this section.

\subsection{Simple examples}\label{SimpleExamples}

{\bf 1.} Suppose one is given the standard sigma model \eqref{S01} and one wants to gauge away ``everything''. For this purpose we choose $E=TM$, the standard Lie algebroid, with $\rho = \mathrm{id}$ and thus $F=M \times 0=: \underline{0}$ in \eqref{FETM}. Now the lift in Eq.~\eqref{FET*M} is performed by the trivial embedding of $TM$ into $TM \oplus T^*M$ ($\alpha = 0$). The only non-trivial equation to satisfy is Eq.~\eqref{Sym}. Since $\bar{\rho}=g$ here, it is sufficient to equip $E=TM$ with any metrical connection (for example $\nabla := \nabla^{Levi-Civita}$) so as to satisfy this equation. The gauging is strict in this case according to the exactness criterium of Theorem \ref{Thmmain1}. On the gauge-theory side, there is an $X^* TM$-valued 1-form gauge field $A=A^i \partial_i$, which equals $V$, while $W=0$, cf.~Eq.~\eqref{VW}. According to Eqs.~\eqref{pullback} and \eqref{Suniv}, 
the action functional now simply becomes
\begin{equation} \label{TM}
S_{gauged}[X,A] = \frac{1}{2}\int_\Sigma g_{ij} (\dd X^i - A^i) \wedge \ast  (\dd X^j - A^j) \, .
\end{equation}
Variation with respect to $A$ gives $\dd X^i = A^i$. With this it is obvious that the $X$-variation does not yield any additional equation. We in particular note that the one equation we have fixes $A$ uniquely in terms of $X$. Knowing that we can deform $X$  arbitrarily, $\delta_\epsilon X^i = \epsilon^i$, we find that the gauge equivalence classes of solutions to the Euler Lagrange equations is not just one point as one might have thought naively. What is true is that this solution space is discrete (zero-dimensional in particular); it consists of the homotopy classes $[X]$ of maps $X \colon\Sigma \to M$. (One may reduce this space further by replacing $\cG_{\cF,0}$ in \eqref{GF} by some bigger subgroup of $\cG_{\cF}$, thus including also ``large gauge transformations'' in the theory).

{\bf 2.} This example is readily generalized in the following way: Let $B \in \Omega^2(M)$ be arbitrary and $H=\dd B$. Choose for the sequence \eqref{FETM} again $\underline{0} \to TM \stackrel{\mathrm{id}}{\longrightarrow} TM$ and the map from $TM \to TM \oplus T^*M, v \mapsto v \oplus \iota_vB$ (which is the graph of the map $B \colon TM \to T^*M, v \mapsto \iota_v B$). Now Eqs.~\eqref{pullback} and \eqref{Suniv} yield \eqref{TM} plus a similar term with $B$, minimally coupled. The conclusion is absolutely unchanged. 

Note that in the previous two examples we discussed some simple (full) Dirac structures, projectable to $TM$. For $B=0$, the second example reduces to the first one, for $B$ closed, $M$ is a pre-symplectic manifold, if furthermore non-degenerate, symplectic. In general, the second example describes an $H$-twisted presymplectic structure. Dirac structures were created to cover presymplectic and Poisson manifolds below the same roof and for $H\neq 0$ their $H$-twisted generalizations \cite{Severa-Weinstein}. We thus now turn to Poisson structures, Dirac structures projectable to $T^*M$:

{\bf 3.} Let $\Pi$ be a Poisson bivector, $[\Pi,\Pi]=0$, and consider the corresponding Lie algebroid $E = T^*M$, cf., e.g., \cite{daSilva-Weinstein99,CLGbook}. Here the anchor map $\rho$ is defined by contraction with $\Pi$, $\rho = \Pi^\sharp \colon T^*M \to TM, \omega \mapsto \iota_\omega \Pi$. Consider first the case of a regular Poisson manifold, i.e.

{\bf 3.1} Let $\Pi$ be of constant rank $R < n$. Then we need a non-trivial bundle $F$ of rank $s=n-R$: $F \to T^*M \to TM$. Take the lift $\sigma$ to be given by the graph of $\Pi^\sharp$, $\sigma \colon T^*M \to TM \oplus T^*M, \omega \mapsto  \iota_\omega \Pi \oplus \omega$, i.e.~we choose $\alpha$ as the identity map. This choice is the one of a  (full) Dirac structure $D \equiv \mathrm{graph}\Pi^\sharp \subset TM \oplus T^*M$, but the lift is certainly not exact.  Thus the gauging is not strict: Although we want to only gauge out the symplectic leaves generated by $\Pi^\sharp$, we have maximal freezing of the propagation of the string transversal to the leaves and the gauged functional becomes topological. Indeed, $S_{gauged}$ has the form of the Poisson sigma model \cite{Schaller-Strobl,Ikeda} with the addition of a kinetic term:
\begin{eqnarray}
S_{gauged}[X,A] &=& \frac{1}{2}\int_\Sigma g_{ij} (\dd X^i + \Pi^{ik} A_k) \wedge \ast  (\dd X^j + \Pi^{jl} A_l ) + 
\nonumber \\ && \int_\Sigma A_i \wedge \dd X^i + \sfrac{1}{2}\Pi^{ij} A_i \wedge A_j
\end{eqnarray}
It was proven in \cite{DSM} that the first line on the right-hand-side can be deleted without changing the Euler-Lagrange equations. After this step, variation with respect to $A$ gives the field equation $\dd X^i + \Pi^{ij} A_j =0$, which implies in particular that $\Sigma$ has to be mapped into a symplectic leaf $L \subset \cF$ of the symplectic foliation $\cF$ generated by $\Pi$ on $M$. The $X$-variation gives some further, independent equations now. This becomes most transparent in the case that one uses a coordinate system on $M$ in which $\Pi$ takes constant Darboux form, showing that for any fixed $X \colon \Sigma \to L$ one has $A\in \Omega^1(\Sigma,X^* N^*L)$ to be closed; here $N^*L\subset T^*M|_L$ denotes the conormal bundle of $L\subset M$. Using the fact that the gauge symmetries \eqref{depXA} reduce to homotopies of $X$ along the symplectic leaves and that the part of $A$ that takes values in the conormal bundle of $L$ can be changed independently by any closed form, we find on the $X$-sector as solutions modulo gauge transformations the homotopy classes $[X]$ of maps from $\Sigma$ to any $L \in \cF$ and on the part of the $A$-sector that is not fixed completely by the choice of $X$ a twisted de Rham cohomology (cf.~\cite{hep-th/0304252} for more details).
 The space of solutions to the Euler-Lagrange equations modulo gauge equivalence is not a discrete space anymore, but it is still finite-dimensional: this is a manifestation of a topological theory on the classical level.

Let us remark that we in fact did not check  that the conditions Eqs.~\eqref{barrholambda} and \eqref{iotaHlambda} present in Theorem \ref{Thmmain1} can be satisfied. However, the Theorem is an if and only if in this context. Thus, we know that there necessarily exists a pair of connections $\nabla$ and $\nabla + \phi$ on $E=T^*M$ such that these constraints are verified. For completeness we provide their explicit form here also: First note that the $\alpha$ being the identity map implies $\alpha =\dd x^i \partial_i$. Using a holonomic basis $\dd x^i$ in $T^*M$ we thus have that $\alpha_a \sim \alpha^i = \dd x^i$ here. 
In this local frame, the coefficients of the connection and of $\phi$ can be chosen according to (e.g.~$\omega^a_{bi} \sim \omega^j_{ki}$):
\bea
\omega^j_{ik}&=&\Gamma^j_{ik}+g_{il}\Pi^{lm}\phi^j_{mk}~,
\\
\phi^j_{ik}&=&-[(1-g\Pi g\Pi)^{-1}]_i^lg_{lm}\nabla_k\Pi^{mj}~.
\eea
For the general formulas of this kind for every possible (full) Dirac structure $D \subset TM \oplus T^*M$, we refer to the accompanying paper \cite{DSMfourofus}.


{\bf 3.2} Take again $\Pi$ and the sequence \eqref{FETM} as above, denote by $(\ker \Pi^\sharp)^\perp$ the subbundle of $T^*M$ which is orthogonal to the subbundle $\ker \Pi^\sharp$ with respect to the metric $g$. Now consider the lift $\sigma$ given by the projection such that $\alpha|_{\ker \Pi^\sharp} = \mathrm{id}$ and $\alpha |_{(\ker \Pi^\sharp)^\perp} =0$. The image $D'$ of the lift $\sigma \colon T^*M \to TM \oplus T^*M$ is 
again a Dirac structure: In fact, $D' =  T\cF \oplus N^*\cF$, where $T\cF \subset TM$ is the tangent distribution to the foliation $\cF$ generated by $\rho \equiv \Pi^\sharp$ and $N^*\cF \subset T^*M$ its conormal bundle. It is then a straightforward to verify that this provides a Dirac structure (a generalization of this construction can be found in Appendix A of \cite{KSS}). Evidently this is \emph{another} Dirac structure for the same foliation and the same sequence \eqref{FETM}, which here takes the form $\ker \rho \to T^*M \to TM$. The gauging is again maximally non-strict and the resulting theory topological. In an adapted coordinate system where $X^I=const$ characterizes the leaves in the given chart, the functional takes the form:
\begin{eqnarray}\label{S'}
S'_{gauged}[X,A] &=& \frac{1}{2}\int_\Sigma g_{ij} (\dd X^i + \Pi^{ik} A_k) \wedge \ast  (\dd X^j + \Pi^{jl} A_l ) + A_I \wedge \dd X^I \, .
\end{eqnarray}
 Note that $\Pi^{ik}A_k$ does not contain $A_I$ and thus the variation with respect to $A_I$ indeed yields maximal freezing, $\dd X^I =0$.

{\bf 3.3} Let us again consider the above sequence 
\eqref{FETM}, but an exact lift. This can be provided by choosing $\alpha \equiv 0$. The image of $\sigma$ is then $T\cF \subset TM \oplus T^*M$, which is evidently involutive and isotropic. It is thus a small, but not a full Dirac structure. Now the action \eqref{SWZ} takes the simple form 
\begin{eqnarray}\label{S''}
S''_{gauged}[X,A] &=& \frac{1}{2}\int_\Sigma g_{ij} (\dd X^i + \Pi^{ik} A_k) \wedge \ast  (\dd X^j + \Pi^{jl} A_l )\, .
\end{eqnarray}
no longer having a full Dirac structure, we can no longer rely on Proposition \ref{D} and  need to determine the conditions on $g$ such that 
\eqref{S''} really provides a gauging. This is precisely the case if $g^{-1}|_{N^*\cF}$ is invariant along the foliation \cite{Kotov:2014iha}.
Since the lifting is exact, under the above condition on $g$, the gauging is strict. 

{\bf 3.4} One can easily imagine an interpolation between the two situations of maximally non-strict gauging \eqref{S'} and strict gauging \eqref{S''}---essentially by dropping part of the additional terms in \eqref{S'} only. The more one drops, the stronger will be the condition on $g$ for the functional to really provide a gauging of the foliation. If one does not drop all of them, resulting into \eqref{S''}, the gauging will remain non-strict, since the strings propagation is frozen in some directions then.

{\bf 3.5} We again take $\Pi$ as above but but now $E=T\cF$, where $\cF$ is the (regular) symplectic foliation on $M$ generated by $\Pi^\sharp$. The anchor $\rho$ then is an embedding and we have to choose $F=\underline{0}$ for exactness of the sequence \eqref{FETM}. The situation is very similar to the one in item {\bf 3.3} above. In fact the lift can be somewhat tilted in the direction of $N^*\cF$ inside $TM \oplus T^*M$ also for keeping a small Dirac structure. Restricting again to $\alpha=0$ all remains essentially the same,  including the condition on $g$. The only difference now is that we do no longer have $n$ gauge fields $A_i$, but $R < n$ ones $A^\alpha$. In an adapted coordinate system $(X^i)=(X^I,X^\alpha)$, the action then takes the form 
\begin{eqnarray}\label{S''g}
S''_{gauged}[X,A] &=& \frac{1}{2}\int_\Sigma g_{IJ} \dd X^I \wedge 
*\dd X^J + g_{\alpha \beta} (\dd X^\alpha - A^\alpha) \wedge \ast  (\dd X^\beta - A^\beta ) +\nonumber \\ 
&&+ \int_\Sigma g_{I\alpha} \dd X^I \wedge * (\dd X^\alpha -A^\alpha)  \, .
\end{eqnarray}
The essential difference between {\bf 3.3} and {\bf 3.5} is that in the first case, $A_I$ drops out of the action completely, corresponding to the $\lambda$-transformation invariance $A_I \mapsto A_I + \lambda_I$ of \eqref{S''}, and effectively replacing the remaining $\Pi^{\alpha \beta} A_\beta$ by $A^\alpha$ (note here that the matrix $(\Pi^{\alpha \beta})$ is invertible).

One may consider many more examples of different kind and of more sophisticated nature, but we leave this task for later work.

\subsection{Obstruction to strict gauging of the SU(2)-WZW model and its almost strict gauging}

\label{SU2WZW}

We now return to an important example of a string propagating in a space-time governed by a 3-flux $H$, namely the WZW-model \eqref{WZW}. Its 3-form $H$, the Cartan 3-form on a compact, semi-simple group $G$, is an example of a closed but not exact 3-form, cf.~Sec.~\ref{Sec:3.2}; however, like in Sec.~\ref{Sec:3.1}, the foliation we will gauge out is given by a conventional group action. In fact, although the model \eqref{WZW} is evidently rigidly invariant with respect to $G \times G$, coming from a constant left- and right-action of $G$ on itself, only particular subgroups of this can be really gauged out---for example, for the whole symmetry group $G \times G$ there are no $\alpha$'s satisfying simultaneously the three consistency conditions \eqref{dalpha}, \eqref{equiv}, and \eqref{iso}. On the other hand, for any subgroup $K \subset G$, the adjoint action, $g(\sigma) \mapsto h g(\sigma) h^{-1}$, $
 h\in K$, can be gauged. 
The resulting string theory is called the $G/K$-WZW model and its action functional has the form:
\begin{eqnarray}\label{G/H}
S_{G/K}[g,A] &=&  \frac{k\hbar}{8\pi} \int_\Sigma 
\mathrm{tr} \left[\left( \dd g g^{-1} + (1-\mathrm{Ad}_g) A \right)\wedge \ast \left( \dd g g^{-1} + (1-\mathrm{Ad}_g) A \right)\right] +\nonumber\\
&& +\frac{k\hbar}{4\pi} \int_\Sigma 
\mathrm{tr} \left[-(1+\mathrm{Ad}_g) A \wedge  \dd g g^{-1} + A \wedge \mathrm{Ad}_g A \right] + \nonumber \\
&&+ \frac{k\hbar}{12\pi} \int\mathrm{tr}
\left[\dd g g^{-1}\wedge \dd gg^{-1}\wedge \dd g g^{-1}
\right] 
\end{eqnarray}
where the Lie-algebra valued gauged-field is $A \in \Omega^1(\Sigma,\mathfrak{k})$, with $\mathfrak{k}$ the Lie algebra of $K$; certainly for $A=0$, we regain the ungauged, original functional $S_{WZW}$ of Eq.~\eqref{WZW}.

Here we focus on the simplest non-trivial example of these models, namely the case where $G=SU(2)$. Then the only option for gauging is $K=G$. This is the completely gauged or $G/G$ WZW-model for the smallest simple compact group, $G=SU(2)$. We will study this gauged theory---as well as potential alternatives, gauging the same orbits---from the perspective of the present paper and in particular for what concerns strictness of the gauging.

For this purpose, we first identify $SU(2)$ with $S^3 \subset \R^4$ by using the parametrization
\begin{equation}
\label{embedding}
g = 
\hat{x}^4\mathbf{1} + i\hat{x}^3\sigma_3 + i\hat{x}^2\sigma_2 + i\hat{x}^1\sigma_1\;,
\end{equation}
where the $\sigma_j$ denote the Pauli matrices. $g\in SU(2)$ is then tantamount to $\hat{x}^J\hat{x}^J=1$, where for capital indices $J,K,\ldots \in \{1,2,3,4\}$, while $(\hat{x}^J)=(\hat{x}^j,\hat{x}^4)$, and lower case indices $j,k\ldots \in \{1,2,3\}$. Adjoint transformations leave invariant the trace of $g$, and thus correspond to constant level sets of $\hat{x}^4$: the orbits we want to factor out are thus 2-spheres of radius $\sqrt{1-(\hat{x}^4)^2}$---except for the two point-like orbits corresponding to $\{\pm 1\} \subset SU(2)$, i.e.~the North and the South pole on the 3-sphere.

Using \eqref{embedding}, we can now rewrite the kinetic  and the WZW term as follows:
\begin{align}
\textrm{tr}(\dd gg^{-1} \wedge \ast \, \dd gg^{-1}) &=\,-2\delta_{JK}\dd\hat{x}^J\wedge \ast \, \dd\hat{x}^K\;,\\
\textrm{tr}(\dd gg^{-1}\wedge \dd gg^{-1}\wedge \dd gg^{-1})&=\,-2\varepsilon_{JKLM}\hat{x}^J
\dd\hat{x}^K\wedge \dd\hat{x}^L\wedge \dd\hat{x}^M\;.
\end{align}
In order to facilitate comparison with \eqref{SWZ} and the other results from Section \ref{Sec3},  we fix the numerical  prefactors in \eqref{G/H} according to $k\hbar/2\pi\to -1$. Now we can read off the standard choice of the $\alpha$'s from the first term in second line of \eqref{G/H}:
\beq\label{all}
\tfrac{1}{2}(\dd gg^{-1}+g^{-1}\dd g)=\,i\alpha_j\sigma_j \, ,
\eeq 
which more explicitly reads 
\begin{equation}
\begin{pmatrix}
i\hat x^4 \dd \hat x^3 -i\hat x^3\dd \hat x^4 & \hat x^4(\dd \hat x^2 + i\dd \hat x^1) - \dd\hat x^4(\hat x^2 + i\hat x^1) \\
-c.c. & -i\hat x^4\dd\hat x^3+i\hat x^3 \dd \hat x^4 \end{pmatrix} = \begin{pmatrix} i\alpha_3 &  \a_2 + i \a_1\\-\a_2+i\a_1 & -i\alpha_3 \end{pmatrix} \,.
\end{equation}
{}From this  we now obtain directly the standard choice of the $\a$s in the SU(2) G/G WZW model:\begin{equation} \label{alphahat}
\alpha_j^{G/G}=\hat x^4\dd \hat x^j-\hat x^j\dd \hat x^4
\end{equation}
The $\gamma$s follow from the $\alpha$s due to Prop.~\ref{Prop3}; it is thus sufficient to focus on the $\alpha_j$ here.

Eq.~\eqref{alphahat} still uses the embedding coordinates $\hat{x}^J$. To obtain genuine charts on the 3-sphere $SU(2)$, we use a stereographic projection; projecting from the North pole is achieved by means of
\begin{equation} 
\hat{x}^i =\frac{2x^i}{r^2+1}~,\quad \hat{x}^4=\,\frac{r^2-1}{r^2+1}~,
\end{equation}
where $r^2=\sum_{i=1}^3\,(x^i)^2$ is the square of the radius for the coordinates $x^i$ on the three-dimensional flat space obtained as the chart covering $SU(2)\backslash \{-1\}$. In this chart, called the N-chart in what follows,
\eqref{alphahat} takes the form
\beq\label{allN}
\alpha_i^{G/G}(x^i)=\frac{4}{(r^2+1)^2}\left(r^2\dd x^i-x^ix^j\dd x^j\right)-\frac{2}{r^2+1}\dd x^i \, ,
\eeq
while the background fields become: 
\bea
\label{gP}
&&g=\frac{2}{(r^2+1)^2}\dd x^i\vee\dd x^i~,\\
\label{HP}
&&H=\frac{16}{(r^2+1)^{3}} \dd x^1\wedge \dd x^2\wedge \dd x^3~.
\eea
Stereographic projection from the South pole is obtained in a similar way, or, by changing coordinates from $x^i$ to $\bar{x}^i$ according to
\begin{equation} \label{coordinatechange}
x^i =\frac{\bar{x}^i}{\bar{r}^2}~,\qquad r^2 =\frac{1}{\bar{r}^2}\;.
\end{equation}
For example, for \eqref{allN} this yields 
\begin{equation}
\label{allS}
\alpha_i^{G/G}(\bar{x}^i)=-\frac{4}{(\bar{r}^2+1)^2}\left(\bar{r}^2\dd \bar{x}^i-\bar{x}^i\bar{x}^j\dd \bar{x}^j\right)+\frac{2}{\bar{r}^2+1}\dd \bar{x}^i \, ,
\end{equation}
which differs in this chart, the S-chart, by an overall minus sign from the expression in the N-chart. Certainly all these expression are well-defined in either chart as they originate from globally well-defined
objects on $SU(2)$.

The stereographic projection has the advantage that now we can take recourse to the analysis of Sec.~\ref{Sec:3.1}. In particular, in each of the charts, the sequence \eqref{FETM} takes the simple form \eqref{FETMso3} with $M = \R^3 \cong S^3\backslash \{-1\}$ for the N-chart and $M = \R^3 \cong S^3\backslash \{1\}$ for the S-chart. In particular, the orbits generated by the adjoint action of $G$ on itself are just concentric 2-spheres around the origin in each of the charts. 

We can now use Prop.~\ref{PropHullstrict} to see if the standard way of gauging of the adjoint action of the SU(2)-WZW model, as given in \eqref{G/H}, is a strict gauging. For this we need to check, if $x^i \alpha_i^{G/G}$ vanishes everywhere on $G=SU(2)$. This is not the case: we obtain
\begin{equation}\label{nonzero}
x^i\alpha_i^{G/G}= \frac{-2}{r^2+1} x^i \dd x^i \neq 0 \, .
\end{equation}
So, as announced already in Sec.~\ref{Orientation}, the degrees of freedom in the $G/G$-model are completely frozen and the gauging is non-strict\footnote{A similar observation was made in Ref.\cite{Bakas:2016nxt} in the context of the  G/G WZW model.}. 

\begin{figure}[ht]
    \centering
  \includegraphics[width=0.7\textwidth]{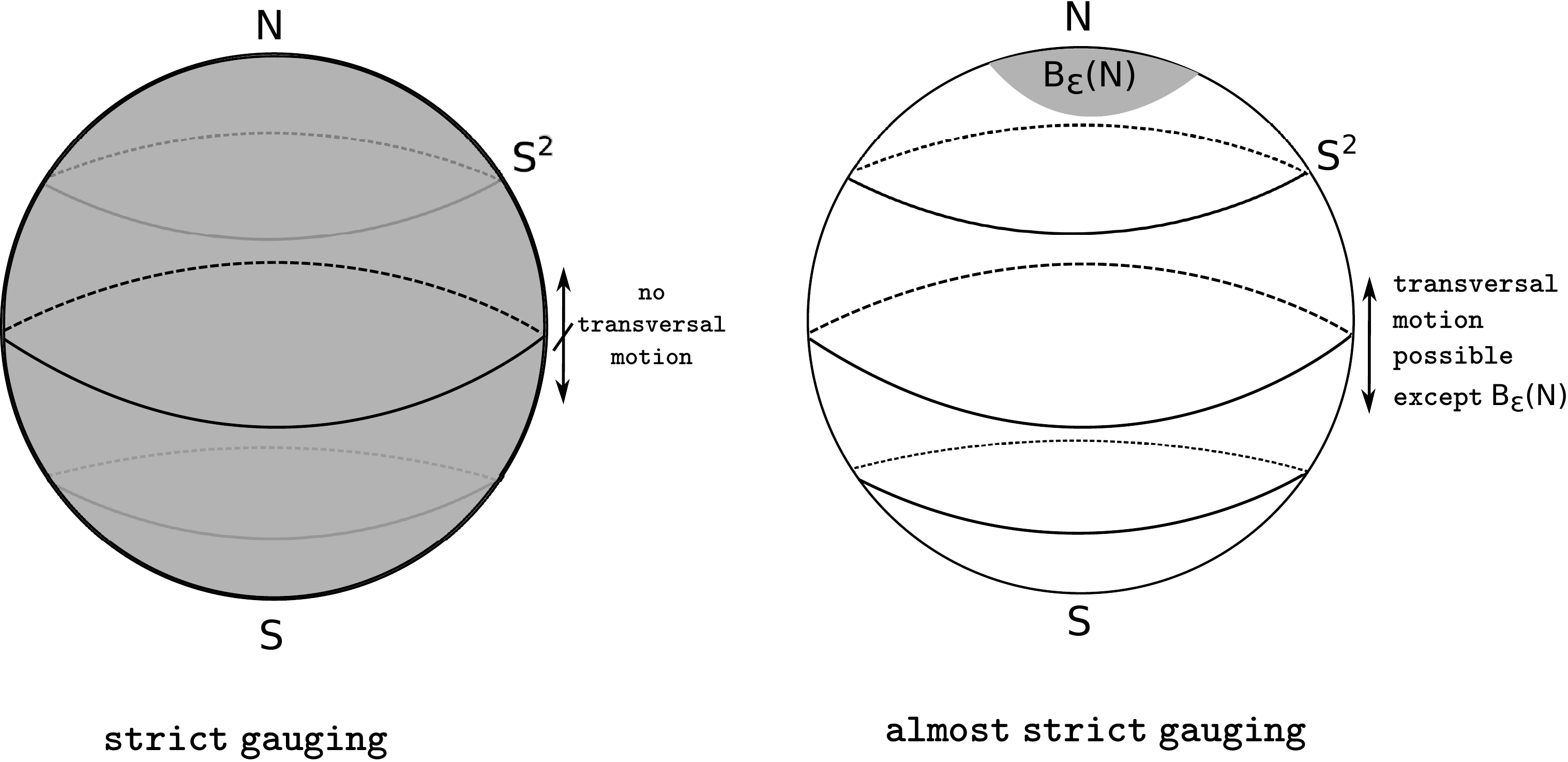}
    \caption{Strict versus almost strict: The total space $M=SU(2)$ is indicated by a circle on both sides, the leaves of the foliation, 2-spheres degenerating to a point at the North and the South pole, by horizontal lines. 
The left-hand picture illustrates the conventional gauging of this situation: a vertical movement of the string, transversal to the orbits of the gauge transformations, is prohibited due to the field equations of the gauged model. The right-hand picture shows a different gauging of the same ungauged WZW-model where now this freezing of the vertical movement is restricted to an arbitrarily small volume around the North pole. Outside of it, the string is no longer constrained and can move freely.}
 \label{figS3} 
\end{figure}

There is now an immediate question posing itself: Is there an alternative choice of the gauging, i.e.~of the choice of $\alpha$s, such that the gauging becomes strict? As we will see, this is obstructed. Then there is a follow-up question: As we see in \eqref{nonzero}, the violation of \eqref{talpha} is maximal on $G=SU(2)$, only on two points, the North and the South pole, this identity is satisfied, i.e.~on a sub-manifold of measure zero. Can we find other choices of $\alpha_i$ such that \eqref{talpha} is satisfied almost everywhere or at least within a region much bigger than where it is violated? Note that we are given a metric on $M$ by the kinetic term, thus we have a notion of volume at hand. Let us formalize this idea as follows:
\begin{defn} 
\label{almost} Let $V$ denote the total volume of the (compact) target $(M,g)$. Let $(B_\epsilon)_{\epsilon >0}$ be a sequence of regions in $M$, $B_\epsilon \subset B_{\epsilon'} \subset M$ for $\epsilon < \epsilon'$, of volume $V_\epsilon$ such that $\lim_{\epsilon \to 0} V_\epsilon/V = 0$.
We call a family $S_{1,\epsilon}$ of gaugings of an ungauged theory $S_0$ \emph{almost strict}, if the gauging is strict on $M\backslash B_\epsilon$ for any choice of $\epsilon>0$.  
\end{defn}
The main result of this analysis is then summarized in:
\begin{prop}
For the WZW-model \eqref{WZW} with $G=SU(2)$, strict gauging of the foliation generated by the adjoint action of $G$ onto itself is obstructed. The standard choice of the gauged theory is maximally non-strict. Almost strict gauging can be achieved, however.
\end{prop}
We summarize these statements in Fig.~\ref{figS3}.

{\bf Proof:} We can use the action Lie algebroid $E=M \times su(2)$. The metric $g$ is strictly invariant w.r.t.~the group action and the natural generators (of a covariantly constant basis) are 2-Hamiltonian w.r.t.~$H$ (cf.~Eq.~\eqref{2symp} below). Thus the two connections $\nabla^\pm = \nabla \pm \phi$ vanish in a covariantly constant basis (a basis of the Lie algebra) and the problem, cf.~Proposition \ref{Prop3}, reduces to the usual one where one needs to find $\alpha$s such that
\begin{equation}\label{2symp}
\iota_{\rho_a} H = \dd \alpha_a
\end{equation}
together with the relations \eqref{equiv} and  \eqref{iso}.
\begin{lemma}\label{uniqueness}
If it exists, strict gauging of the given foliation on $M=S^3$ (or on any rotation-invariant connected region containing the North pole or South pole) is unique.  
\end{lemma}
{\bf Proof} {\it (of Lemma \ref{uniqueness})}{\bf :}
Let us restrict to the N-chart first. 
Then $\alpha$'s satisfying the condition \eqref{uniqueness} are unique up to the addition of one-forms $\dd f_a$ for $f_a \in {\cal C}^{\infty}(\R^3)$. Strictness now reads $x^a\alpha_a=0$. We want to show that, together with the other conditions to be satisfied by $\alpha_a$, this will imply $f_a = 0$ for all $a=1,2,3$, i.e.~the required uniqueness. First, from $x^a\dd f_a=0$, we have
$$0= \dd(f_a \dd x^a)\;,\quad \textrm{i.e.}\quad f_a \dd x^a = \dd F~,$$
for a function $F$. This means that $f_a = \partial_a F$. As a next step, from the rotation equivariance of $\dd f_a$ (as the $\alpha_a$ have to be rotation equivariant 1-forms), it follows that $F$ is a function which is rotation \emph{invariant}. Returning to $x^a\dd(\partial_a F) = 0$, we then get 
$$x^a\dd(\partial_a F) = \dd(x^a\partial_a F) - \partial_a F \dd x^a = \dd(rF'(r) - F(r))=0~, $$
which yields $F(r)=ar$ for some constant $a$. This, in turn, implies however, that $f_a=a x^a/r$, which is singular at the North pole for $a\neq 0$. We therefore conclude $f_a=0$. $\square$ {\it (Lemma)}

However, such an $\alpha_a$ cannot exist. To show this, we again return first to the N-chart. This will then also provide a way of achieving almost strictness. Restricting to the N-chart, we can apply Lemma \ref{lemmaBinv} so as to find a and then by Lemma \ref{uniqueness} the strict choice of $\alpha_a$ provided we find an invariant primitive $B$ for $H$, $H|_{\R^3}= \dd B$. Using the expression for   $H$ in the N-chart given in \eqref{HP}, one verifies that 
\bea
&&B^{inv}_{N}=h(r)\epsilon_{abc}x^a\dd x^b\w \dd x^c~,\label{BN}\\
&&h(r)=\frac{r^2-1}{r^2(r^2+1)^2}+\frac{1}{r^3}\arctan r, \nn 
\eea
is such an invariant primitive. It is important to note that this expression is well-defined at the origin, which represents the North pole of $S^3$, as the divergencies of the two terms in $h$ cancel against one another. Thus, according to Lemma \ref{lemmaBinv}, there exists strict gauing in all of the N-chart, given by  $\left(\alpha^{\textrm{strict}}_N\right)_a=-
\iota_{\rho_a}B^{inv}_{N}$. Using $\rho_a = \epsilon_{abc} x^a \partial_b$, this yields
\beq
\label{anorth}
\left(\alpha^{\textrm{strict}}_N\right)_a =2 h(r)\left( r^2\dd x^a -x^ax^b\dd x^b\right) \equiv : \gamma_a (x) ~,
\eeq
where the abbreviation $\gamma_a$ is introduced for later use.

To see, if this expression extends also to the South pole when transformed into the S-chart, we can apply the  transformation (\ref{coordinatechange}). This yields for the coordinates in the S-chart
\begin{equation} \left(\alpha_N^{\textrm{strict}}\right)_{a}|_{N \cap S}(\bar r)  =2\frac{h(1/\bar r)}{\bar r^6}\left( \bar r^2 \dd\bar x^a -\bar x^a\bar x^b\dd\bar x^b\right)~.
\end{equation}
Using
\beq\label{hsouth}
h(1/\bar r)=-\bar r^6\left(\frac{\bar r^2-1}{\bar r^2(\bar r^2+1)^2}+\frac{1}{\bar r^3}\arctan(\bar r)-\frac{\pi}{2\bar r^3}\right)~,
\eeq
we may rewrite this expression as 
\beq\label{rew}
\left(\alpha_N^{strict}\right)_a = -\gamma_a(\bar x)+\dd\left(\pi\frac{\bar x^i}{\bar r}\right),
\eeq
where the formula for the 1-forms $\gamma_a$ have been provided in (\ref{anorth}). While these expressions are well-defined for $\bar{r}\to 0$, the additional exact 1-forms in \eqref{rew} are evidently not. Together with the uniqueness of the 1-forms giving strict gauging on the N-chart, this proves non-existence of a strict gauging on all of $SU(2) \cong S^3$.   

On the other hand, these expressions show how to obtain an almost strict gauging of the WZW model for the adjoint orbits on $SU(2)$. For this purpose we first choose a compactly supported smooth function $F_\epsilon$ around the South pole such that
\begin{equation}
 F_\epsilon(\bar r) =1, \;\textrm{for}\; \bar{r}\geq  \bar R_\epsilon \, \qquad F_\epsilon(\bar r) = 0, \; \textrm{for}\;0\leq \bar r \leq  \frac{\bar{R}_\epsilon}{2}
\end{equation}
for some null sequence $(\bar{R}_\epsilon)_{\epsilon >0}$. Now define
$\left(\alpha^{astrict}_\epsilon\right)_a$ as follows: In the N-chart, one has
\beq \label{alphaN}
\left(\alpha^{astrict}_\epsilon\right)_a |_N := \gamma_a(x) \, , \quad \textrm{for}\; r \leq \frac{1}{2\bar{R}_\epsilon}
\eeq
and on the S-chart
\beq \label{alphaS}
\left(\alpha^{astrict}_\epsilon\right)_a |_S := -\gamma_a(\bar x)+\dd\left(\pi F_\epsilon(\bar{r})\frac{\bar x^i}{\bar r}\right)
\eeq
This now glues together smoothly to  1-forms $\alpha_a^{astrict}$ that are globally defined on $S^3$. They now violate the strictness condition only for distances determined by $\bar{r}<\bar{R}_\epsilon$ and thus within a region around the South pole of a volume $V_\epsilon$ which tends to zero with $\epsilon$. \hfill $\square$

We conclude this subsection with some remarks: 

\begin{itemize}
\item Since 
\begin{equation}
\dd \left(\frac{x^a}{r}\right) = \tfrac{1}{r^3}\Bigl(r^2dx^a - x^a x_b dx^b\Bigr) \, ,
\end{equation} 
which vanishes identically upon contraction with $x^a$, and $\gamma_a(x) x^a \equiv 0$, $\alpha^{astrict}_\epsilon$ does not violate the strictness condition for both $r > \bar R_\epsilon$ and $\bar{r}<\frac{\bar R_\epsilon}{2}$, so with the gauged action \eqref{SWZ} using this $\alpha$, the transversal movement of the string can be frozen only within a shell around the South pole. 
\item As we saw above, there is a strict gauging when restricting the target to $SU(2)\backslash \{-1\} \cong \R^3$ which is determined by means of \eqref{anorth}. One verifies that this strict gauging differs from the standard one by the addition of exact 1-forms,
\begin{equation}
(\alpha_a^{G/G})|_N - (\alpha_N^{\textrm{strict}})_a =\dd f_a\;,\quad f_a =\, \frac{x^a}{r} \arctan(r)\;,
\end{equation}
as it has to be. Certainly, the functions $f_a$ do not have a unique limit for $r \to \infty$ and, as we saw above, neither do their differentials. 
\item Since $H$ has a non-trivial cohomology class, 
its primitive $B_N^{inv}$ has to diverge when extended to the South pole. Indeed, transforming \eqref{BN} by means of \eqref{coordinatechange}, one finds 
\begin{equation}\label{Bneu}
B^{inv}_N |_{N\cap S}=\frac{h(1/\bar r)}{\bar r^6}\epsilon_{abc}\bar x^a\dd\bar x^b\w \dd\bar x^c \, .
\end{equation}
Using the behaviour of $h(1/\bar{r})$ given in \eqref{hsouth}, we see that its last contribution indeed renders \eqref{Bneu} singular at the South pole.
\item Certainly, one can find a similar invariant potential for $H$ defined in all of the S-chart,
 \begin{equation}
B^{inv}_S =-h(\bar r)\epsilon_{abc}\bar x^a\dd\bar x^b\w \dd\bar x^c \, ,
\end{equation}
and use it for the construction of $(\alpha_S^{strict})_a$. These $\alpha$s then cannot be extended to the North pole, but by additions of  exact contributions around the identity element can be transformed into another almost strict gauging. It is the gauged theory for this latter choice of $\alpha$s that is depicted in Fig.~\ref{figS3}.
\item There is one more non-trivial consistency check that one may perform: $\gamma_{ab}$ is to be calculated as a contraction of $\rho_a$ with  $\alpha_b^{astrict}$. If one does this in the N-chart, one obtains from \eqref{alphaN} after a calculation \begin{equation}
\gamma_{ab}|_N = -2 h(r)r^2 \epsilon_{abc} x^c \qquad \mathrm{for} \quad f \leq \frac{1}{2\bar{R_\epsilon}}
\end{equation} 
On the other hand, this has to agree with $\bar{\rho_a}$ contracted with \eqref{alphaS} when transformed back into the N-chart. One may verify that this is indeed the case. On the way of doing so, one uses $\rho_a\equiv \epsilon_{abc} x^b \partial_c = \epsilon_{abc} \bar{x}^b \partial_{\bar{c}}\equiv \bar{\rho_a}$, following directly from \eqref{coordinatechange}, as well as the property \eqref{hsouth} of $h$. 

\end{itemize}

\paragraph{Acknowledgements.} 
T.S.~is grateful to Henrique Bursztyn, Gil Calvacanti, Chris Hull, Alexei Kotov and  Ryszard Nest for valuable discussions.  He is particularly grateful to  Camille Laurent-Gengoux for essential discussions in the context of  \cite{LG-Lavau-Strobl}, 
which also played an important role for the setting of this paper. A.C. would like to thank Vladimir Salnikov for assistance with the figures. A.D. is grateful to Christian Saemann, David Berman and Ralph Blumenhagen for discussion. We are also very grateful to Jim Stasheff for valuable 
editorial comments. 

The work of L.J.~was supported by the Croatian Science Foundation under the project IP-2014-09-3258, as
well as by the H2020 Twinning project No. 692194, ``RBI-T-WINNING''.
The work of T.S.~was partially supported by Projeto P.V.E. 88881.030367/
2013-01 (CAPES/Brazil), Anton Alekseev's project MODFLAT of the European Research Council
(ERC), and the NCCR SwissMAP of the Swiss National Science Foundation.


\appendix

\section{Most General Ansatz of Gauged Strings with H-flux}
\label{AppendixA}

In this appendix we consider adding more terms to the Ansatz for to the action functional with WZ-term and its gauge transformations. In particular we now consider the action functional
\begin{equation} \label{ansatzgen}
S_{2WZ}[X,A]=\int_\Sigma \sfrac{1}{2}g_{ij} \dd X^i \w \ast \dd X^j + \int H + \int_\Sigma A^a \wedge \alpha_a +
A^a\w\ast \tilde\alpha_a+ \sfrac{1}{2} \gamma_{ab} A^a \wedge A^b
+\sfrac 12 \tilde\g_{ab}A^a\w\ast A^b \: , 
\end{equation}
where $\alpha_a$ and $\tilde\alpha_a$ are 1-forms on $M$, and $\gamma_{ab}$ and $\tilde\gamma_{ab}$ functions, all pulled back by $X$ to $\Sigma$. There are 
two additional terms here in comparison to the main text. 
Both of these terms contain a Hodge operator. This means 
that much like the WZ-term where general gauging requires 
to abandon minimal coupling, thus introducing 
non-minimal \emph{topological} terms with $A^a$, we allow 
now for the possibility that the non-minimally coupled gauge 
field induces also additional terms in the \emph{kinetic} 
sector.
Note that in principle we could have also added terms proportional to $\dd A^a$ and $\ast\dd A^a$; however, 
such terms may always be absorbed in the already present ones. 
As for the infinitesimal gauge transformations, we consider similarly enlarging our Ansatz from the main text considerably:
\bea 
\label{deltaX2} \d_{\epsilon}X^i&=&v^i_a(X)\epsilon^a~,\\
\label{deltaA2}
\d_{\epsilon} A^a&=&r^a_b(X)\dd\epsilon^b+s^a_b(X)\ast\dd\epsilon^b+C^a_{bc}(X)A^b\epsilon^c+ \omega^a_{bi}(X)\epsilon^b \, \dD X^i + \phi^a_{bi}(X) \epsilon^b \ast \!\dD X^i+\nn\\
&+&\chi^a_{bc}(X)A^b
\epsilon^c+\psi^a_{bc}(X)\ast A^b\epsilon^c~.
\eea
The gauge transformation of the scalar fields $X^i$ remains the same as in the body of the text due to the second main requirement that was described in the introduction. 

The next step is to compute the gauge variation of the action functional. Clearly this once more leads to a set of 
extended invariance conditions for $g$ and $H$ accompanied by a set of constraints that obstruct the gauging. We are not going to present a step-by-step calculation here; instead we state the final results. The conditions for $g$ and $H$ now become
\bea \label{gencondition1}
{\cal L}_{v_a}g&=&-\omega^b_a\vee\tilde \alpha_b
+\phi_a^b\vee\alpha_b~,\\
\label{gencondition2}
\iota_{v_a}H&=&\dd(r^b_a\alpha_b)\pm 
\dd(s_a^b\tilde\alpha_b)-\omega^b_a\w\alpha_b
\mp\phi^b_a\w\tilde\alpha_b~.
\eea 
Before we proceed with the rest of the constraints let us pause for a moment to justify that these conditions reduce to the ones found in the body of the paper for the appropriate special case: 
The functional and gauge transformations of Section \ref{Sec3} are reproduced from the above by means of choosing 
\begin{equation} \label{mainansatz} r^b_a=\d^b_a, \quad s^b_a=\chi^c_{ab}=
\psi^c_{ab}=0, \quad \tilde \alpha_a=-\iota_{v_a}g, \quad 
\tilde\gamma_{ab}= \iota_{v_a}\iota_{v_b}g\,.
\end{equation}
Indeed, one finds that for this choice the Equations \eqref{gencondition1} and \eqref{gencondition2} become \eqref{condition1gH} and 
\eqref{condition2gH}, respectively.

As before, this is not the end of the story, since there are  additional constraints that obstruct the gauging. These are found to be
\bea 
\label{constraint1}
 r_a^c\g_{cb}\pm  s_a^c\tilde\g_{cb}&=&\iota_{v_a}\alpha_b~,
 \\
\label{constraint2}
 s_a^c\g_{cb}-r_a^c\tilde\g_{cb}&=&\iota_{v_a}\tilde\a_b~,\\
\label{constraint5}
 s^b_a\a_b-r^b_a\tilde\a_b&=&\iota_{v_a}g~,
\\ 
 {\cal L}_{v_a}\alpha_b&=&(C^c_{ab}-\chi^c_{ba})\alpha_c
\mp\psi^c_{ba}\tilde\a_c+\iota_{v_b}\dd(r_a^c\a_c\pm s_a^c\tilde\a_c)+\iota_{v_a}\iota_{v_b}H+\nn\\
&&  + ~\omega^c_a(\gamma_{cb}+\iota_{v_b}\alpha_c)
\pm\phi^c_a(\tilde\g_{cb}+\iota_{v_b}\tilde\a_c)~,
\label{constraint3}
\\
 {\cal L}_{v_a}\tilde\a_b
&=&(C^c_{ab}-\chi^c_{ba})\tilde \a_c+ \psi^c_{ba}\alpha_c-\iota_{v_b}{\cal L}_{v_{a}}g - \nn\\ 
&& -\omega^c_a(\tilde\g_{cb}
+\iota_{v_b}\tilde\a_c)
+\phi^c_a(\g_{cb}+\iota_{v_b}\a_c)~,
\label{constraint4}
\\
\sfrac 12 {\cal L}_{v_a}\g_{cb}&=&\g_{d[c}(C^d_{b]a}+\chi^d_{b]a})\pm \tilde \g_{d[c}\psi^d_{b]a}-\g_{d[c}\iota_{v_{b]}}\omega^d_a
\mp \tilde \g_{d[c}\iota_{v_{b]}}\phi^d_a~,\label{constraint6}
\\
\sfrac 12 {\cal L}_{v_a}\tilde\g_{cb}&=&-\tilde\g_{d[c}(C^d_{b]a}+\chi^d_{b]a})+
\g_{d[c}\psi^d_{b]a}+\tilde\g_{d[c}\iota_{v_{b]}}\omega^d_a
- \g_{d[c}\iota_{v_{b]}}\phi^d_a~.\label{constraint7}
\eea
Once more one may perform the consistency check that these equations reduce to the constraints found in the main text
for the special choice made in Eq.~\eqref{mainansatz}. 
Then Eq.~\eqref{constraint1} reduces to $\g_{ab}=\iota_{v_a}\a_b$ of Proposition \ref{Prop3}
and the successive two constraints \eqref{constraint2} and \eqref{constraint5} become vacuous now. Furthermore, 
Eq. \eqref{constraint3} becomes identical to Eq. \eqref{Calpha} of Proposition \ref{Prop3} and 
Eq. \eqref{constraint4} follows from the rest. Finally, as regards the last two constraints, Eqs. \eqref{constraint6} and 
\eqref{constraint7}, they both follow from the rest in this limit. Thus we verify that with the choices appearing in 
Eq.~\eqref{mainansatz} there are only two obstructing constraints, indeed the ones presented in the main text.

In the present paper we will not investigate the apparently more general situation which seems to enrol itself here any further. We provide the formulas here still for completeness. We also remark that the Equations \eqref{gencondition1} and \eqref{gencondition2} are more symmetrical now than the corresponding equations of the main text, Eqs.~\eqref{condition1gH} and 
\eqref{condition2gH} or, in frame-independent form, Eqs.~\eqref{barrholambda} and \eqref{iotaHlambda}. It thus would be interesting to study the geometric significance as well as the corresponding gauge theory in more detail elsewhere.

\bibliography{DSM}  

\providecommand{\href}[2]{#2}\begingroup\raggedright\begin{thebibliography}{10}

\bibitem{Polyakov}
A.~M. Polyakov, ``{Quantum Geometry of Bosonic Strings},'' {\em Phys. Lett.}
  {\bf B103} (1981) 207--210.

\bibitem{Green}
M.~B. Green, J.~H. Schwarz, and E.~Witten, {\em {Superstring Theory. Vol. 1:
  Introduction}}.
\newblock Cambridge Monogr. Math. Phys., 1988.

\bibitem{Polchinski}
J.~Polchinski, {\em {String theory. Vol. 1: An introduction to the bosonic
  string}}.
\newblock Cambridge University Press, 2007.

\bibitem{orb1}
L.~J. Dixon, J.~A. Harvey, C.~Vafa, and E.~Witten, ``{Strings on Orbifolds},''
  {\em Nucl. Phys.} {\bf B261} (1985) 678--686.

\bibitem{orb2}
L.~J. Dixon, J.~A. Harvey, C.~Vafa, and E.~Witten, ``{Strings on Orbifolds.
  2.},'' {\em Nucl. Phys.} {\bf B274} (1986) 285--314.

\bibitem{WZW}
E.~Witten, ``{Nonabelian Bosonization in Two-Dimensions},'' {\em Commun. Math.
  Phys.} {\bf 92} (1984) 455--472.

\bibitem{Hull2}
C.~M. Hull and B.~J. Spence, ``{The Gauged Nonlinear $\sigma$ Model With
  {Wess-Zumino} Term},'' {\em Phys. Lett.} {\bf B232} (1989) 204.

\bibitem{Figueroa}
J.~M. Figueroa-O'Farrill and S.~Stanciu, ``{Equivariant cohomology and gauged
  bosonic sigma models},'' \href{http://xxx.lanl.gov/abs/hep-th/9407149}{{\tt
  hep-th/9407149}}.

\bibitem{Figueroa2}
J.~M. Figueroa-O'Farrill and S.~Stanciu, ``{Gauged Wess-Zumino terms and
  equivariant cohomology},'' {\em Phys. Lett.} {\bf B341} (1994) 153--159,
  \href{http://xxx.lanl.gov/abs/hep-th/9407196}{{\tt hep-th/9407196}}.

\bibitem{Alekseev-Strobl}
A.~Alekseev and T.~Strobl, ``{Current algebras and differential geometry},''
  {\em JHEP} {\bf 03} (2005) 035,
  \href{http://xxx.lanl.gov/abs/hep-th/0410183}{{\tt hep-th/0410183}}.

\bibitem{Gawedzki1}
K.~Gawedzki and A.~Kupiainen, ``{{G/H} Conformal Field Theory from Gauged WZW
  Model},'' {\em Phys. Lett.} {\bf B215} (1988) 119--123.

\bibitem{Gawedzki2}
K.~Gawedzki and A.~Kupiainen, ``{Coset Construction from Functional
  Integrals},'' {\em Nucl. Phys.} {\bf B320} (1989) 625.

\bibitem{Stroblinprep}
T.~Strobl. in preparation.

\bibitem{LG-Lavau-Strobl}
S.~L. Camille Laurent-Gengoux and T.~Strobl. in preparation.

\bibitem{Gruetzmann-Strobl}
M.~Gr\"utzmann and T.~Strobl, ``General {Y}ang-{M}ills type gauge theories for
  p-form gauge fields: {F}rom physics-based ideas to a mathematical framework
  or {F}rom {B}ianchi identities to twisted {C}ourant algebroids,'' {\em
  International Journal of Geom. Methods in Modern Physics} {\bf 12} (2015),
  no.~01 1550009, \href{http://xxx.lanl.gov/abs/1407.6759}{{\tt 1407.6759}}.

\bibitem{BKS}
M.~Bojowald, A.~Kotov, and T.~Strobl, ``Lie algebroid morphisms, {P}oisson
  {S}igma {M}odels, and off-shell closed gauge symmetries,'' {\em J.Geom.Phys.}
  {\bf 54} (2005) 400--426, \href{http://xxx.lanl.gov/abs/math/0406445}{{\tt
  math/0406445}}.

\bibitem{DSMfourofus}
A.~Chatzistavrakidis, A.~Deser, L.~Jonke, and T.~Strobl, ``{Beyond the standard
  gauging: gauge symmetries of Dirac Sigma Models},''
  \href{http://xxx.lanl.gov/abs/1607.00342}{{\tt 1607.00342}}.

\bibitem{Killing2}
A.~Kotov and T.~Strobl, ``{Geometry on Lie algebroids I: compatible geometric
  structures on the base},'' \href{http://xxx.lanl.gov/abs/1603.04490}{{\tt
  1603.04490}}.

\bibitem{Hitchin03}
N.~Hitchin, ``{Generalized Calabi-Yau manifolds},'' {\em Quart. J. Math.} {\bf
  54} (2003) 281--308, \href{http://xxx.lanl.gov/abs/math/0209099}{{\tt
  math/0209099}}.

\bibitem{Kotov-Strobl10}
A.~Kotov and T.~Strobl, ``{Generalizing Geometry - Algebroids and Sigma
  Models},'' \href{http://xxx.lanl.gov/abs/1004.0632}{{\tt 1004.0632}}.

\bibitem{ScherkSchwarz}
J.~Scherk and J.~H. Schwarz, ``{How to Get Masses from Extra Dimensions},''
  {\em Nucl. Phys.} {\bf B153} (1979) 61--88.

\bibitem{MaharanaSchwarz}
J.~Maharana and J.~H. Schwarz, ``{Noncompact symmetries in string theory},''
  {\em Nucl. Phys.} {\bf B390} (1993) 3--32,
  \href{http://xxx.lanl.gov/abs/hep-th/9207016}{{\tt hep-th/9207016}}.

\bibitem{Kotov:2014iha}
A.~Kotov and T.~Strobl, ``{Gauging without initial symmetry},'' {\em J. Geom.
  Phys.} {\bf 99} (2016) 184--189,
  \href{http://xxx.lanl.gov/abs/1403.8119}{{\tt 1403.8119}}.

\bibitem{Hull1}
C.~M. Hull, A.~Karlhede, U.~Lindstrom, and M.~Rocek, ``{Nonlinear $\sigma$
  Models and Their Gauging in and Out of Superspace},'' {\em Nucl. Phys.} {\bf
  B266} (1986) 1.

\bibitem{Plauschinn}
E.~Plauschinn, ``{On T-duality transformations for the three-sphere},'' {\em
  Nucl. Phys.} {\bf B893} (2015) 257--286,
  \href{http://xxx.lanl.gov/abs/1408.1715}{{\tt 1408.1715}}.

\bibitem{KSS}
A.~Kotov, V.~Salnikov, and T.~Strobl, ``{2d Gauge Theories and Generalized
  Geometry},'' {\em JHEP} {\bf 08} (2014) 021,
  \href{http://xxx.lanl.gov/abs/1407.5439}{{\tt 1407.5439}}.

\bibitem{Anton1}
A.~Alekseev and E.~Meinrenken, ``The non-commutative weil algebra,'' {\em
  Inventiones mathematicae} {\bf 139} (2000), no.~1 135--172,
  \href{http://xxx.lanl.gov/abs/math/9903052}{{\tt math/9903052}}.

\bibitem{Alekseev-Severa}
A.~Alekseev and P.~{\v S}evera, ``{Equivariant cohomology and current
  algebras},'' \href{http://xxx.lanl.gov/abs/1007.3118}{{\tt 1007.3118}}.

\bibitem{DSM}
A.~Kotov, P.~Schaller, and T.~Strobl, ``{Dirac sigma models},'' {\em Commun.
  Math. Phys.} {\bf 260} (2005) 455--480,
  \href{http://xxx.lanl.gov/abs/hep-th/0411112}{{\tt hep-th/0411112}}.

\bibitem{Severa-Weinstein}
P.~{\v S}evera and A.~Weinstein, ``{Poisson geometry with a 3 form
  background},'' {\em Prog. Theor. Phys. Suppl.} {\bf 144} (2001) 145--154,
  \href{http://xxx.lanl.gov/abs/math/0107133}{{\tt math/0107133}}.

\bibitem{Dorfmann}
I.~Y. Dorfman, ``Dirac structures of integrable evolution equations,'' {\em
  Physics Letters A} {\bf 125} (1987), no.~5 240 -- 246.

\bibitem{Courant}
T.~Courant, ``Dirac manifolds,'' {\em Trans.A.M.S.} {\bf 319} (1990) 631--661.

\bibitem{lwx}
Z.-J. Liu, A.~Weinstein, and P.~Xu, ``Manin triples for lie bialgebroids,''
  {\em J. Differential Geom.} {\bf 45} (1997), no.~3 547--574,
  \href{http://xxx.lanl.gov/abs/dg-ga/9508013}{{\tt dg-ga/9508013}}.

\bibitem{Severa-Letters}
P.~{\v S}evera, ``Letters to {A}lan {W}einstein,''.
  http://sophia.dtp.fmph.uniba.sk/~severa/letters/.

\bibitem{Schaller-Strobl}
P.~Schaller and T.~Strobl, ``Poisson structure induced (topological) field
  theories,'' {\em Mod. Phys. Lett.} {\bf A9} (1994) 3129--3136,
  \href{http://xxx.lanl.gov/abs/hep-th/9405110}{{\tt hep-th/9405110}}.

\bibitem{Ikeda}
N.~Ikeda, ``Two-dimensional gravity and nonlinear gauge theory,'' {\em Ann.
  Phys.} {\bf 235} (1994) 435--464,
  \href{http://xxx.lanl.gov/abs/hep-th/9312059}{{\tt hep-th/9312059}}.

\bibitem{daSilva-Weinstein99}
A.~da~Silva and A.~Weinstein, {\em {Geometric Models for Noncommutative
  Algebras}}.
\newblock Berkeley Mathematics Lecture Notes {\bf 10}, 1999.

\bibitem{CLGbook}
C.~Laurent-Gengoux, A.~Pichereau, and P.~Vanhaecke, {\em Poisson Structures}.
\newblock Springer Berlin Heidelberg, Berlin, Heidelberg, 2013.

\bibitem{hep-th/0304252}
M.~Bojowald and T.~Strobl, ``{Classical solutions for Poisson sigma models on a
  Riemann surface},'' {\em JHEP} {\bf 07} (2003) 002,
  \href{http://xxx.lanl.gov/abs/hep-th/0304252}{{\tt hep-th/0304252}}.

\bibitem{Bakas:2016nxt}
I.~Bakas, D.~L{\"u}st, and E.~Plauschinn, ``{Towards a world-sheet description
  of doubled geometry in string theory},''
  \href{http://xxx.lanl.gov/abs/1602.07705}{{\tt 1602.07705}}.

\end{thebibliography}\endgroup
\bibliographystyle{utphys}

\end{document}